\documentclass[preprint,times,sort&compress]{elsarticle}
\usepackage{amssymb}
\usepackage{amsfonts}
\usepackage{amsmath}
\usepackage{color}
\usepackage{soul}
\usepackage{graphicx}
\usepackage{epstopdf}

\begin{document}

\title{Ni coarsening in the three-phase solid oxide fuel cell anode \--- a phase-field simulation study}
\author[UM]{Hsun-Yi Chen}

\author[UM]{Hui-Chia Yu}%

\author[NW]{J. Scott Cronin}%

\author[NW]{James R. Wilson}%

\author[NW]{Scott A. Barnett}%

\author[UM]{Katsuyo Thornton}
\ead{kthorn@umich.edu}

\address[UM]{Department of Materials Science and Engineering, University of Michigan, 2300 Hayward St., Ann Arbor, MI 48109, USA}
\address[NW]{Department of Materials Science and Engineering, Northwestern University, 2220 Campus Drive, Evanston, IL 60201, USA}
\date{\today}

\begin{abstract}
Ni coarsening in Ni-yttria stabilized zirconia (YSZ) solid oxide fuel cell anodes is considered a major reason for anode degradation.  We present a predictive, quantative modeling framework based on the phase-field approach to systematically examine coarsening kinetics in such anodes.  The initial structures for simulations are experimentally acquired functional layers of anodes.  Sample size effects and error analysis of contact angles are examined.  Three phase boundary (TPB) lengths and Ni surface areas are quantatively identified on the basis of the active, dead-end, and isolated phase clusters  throughout coarsening. Tortuosity evolution of the pores is also investigated. We find that phase clusters with larger characteristic length evolve slower than those with smaller length scales. As a result, coarsening has small positive effects on transport, and impacts less on the active Ni surface area than the total counter part.  TPBs, however, are found to be sensitive to local morphological features and are only indirectly correlated to the evolution kinetics of the Ni phase.  

\end{abstract}


\maketitle

\section{Introduction}
\subsection{Nickel coarsening in SOFC anodes}

Solid oxide fuel cells (SOFCs) are one of the most promising clean energy conversion devices for stationary applications because of their low pollutant emissions, high efficiency, and ability to operate using various hydrocarbon fuels. The need for precious-metal catalysts is eliminated in SOFCs because the reaction kinetics is enhanced at their operating temperatures, which are between 500 and 1000$^\circ$C~\citep{Atkinson}. However, high operating temperatures also lead to disadvantages such as slow startup, high fabrication costs, and rapid component degradation~\citep{SMHaile}.  For SOFCs to be commercially viable for stationary applications, their lifetime must meet a minimum of $\sim$50,000 hours. This goal has not yet been achieved. Understanding SOFC degradation mechanisms is therefore crucial to improving their durability.

The degradation mechanisms of SOFCs have been reviewed in a few articles~\citep{Yokokawa08,TuJPS04}. SOFC degradation is usually evaluated in terms of cell power or voltage decrease or area-specific resistance (ASR) increase, which can be determined by the electrochemical impedance spectroscopy (EIS). These electrochemical methods allow the cell degradation to be monitored in situ. However, even though a few studies~\citep{Hagen1,Hagen2} have attempted to address SOFC degradation, it is very difficult to unambiguously characterize the impacts of individual mechanisms using these techniques. Many non-electrochemical methods have been used to identify and monitor the degradation and failure mechanisms of SOFCs in situ or pre-/post-operation (for a review, see Ref.~\cite{Malzbender09}), including X-ray tomography and X-ray diffraction.  Among these degradation mechanisms, the microstructural change in a Ni-based cermet anode is one of the least understood, because experiments that can provide detailed time-dependent, three-dimensional (3D) structural information are difficult.  

The anode of SOFCs is usually made of a composite material of complex morphology that facilitates electrochemical reactions, which require simultaeous transport of fuels, ions, and electrons.  The most commonly used anode material up to date is a porous cermet comprised of Ni and YSZ. Electrochemical reactions at the SOFC anode mainly occur at the vicinity of TPBs, where the pore, Ni, and YSZ phases are in contact~\citep{Mogensen}.  Although the electrochemical reaction mechanisms are not very well understood, the length of TPBs is considered one of the most important geometrical parameters that dictates the resistance in a SOFC anode~\citep{SSunde}.  The anode is thus designed to have a complex microstructure and increase the TPB lengths and the simultaneous transport of three different species.  However, this intricate anode microstructure is typically not stable. The agglomeration and coarsening of nickel particles have been considered the major mechanisms responsible for microstructural change in SOFC anodes~\citep{Simwonis}. Ni coarsening in the SOFC anode is a capillarity-driven phenomenon.  Regions with high curvatures have higher chemical potentials than those with lower curvatures in accordance with the Gibbs-Thomson effect. Materials will therefore be transported from higher- to lower-curvature regions when mobility is sufficiently large for the time scale of interest. This phenomenon increases the resistance in SOFCs and results in cell degradation.

Long-term coarsening experiments have been conducted for thousands of hours to study Ni coarsening in Ni-YSZ anodes.  Simwonis et al.~\citep{Simwonis} measured a $33\%$ decrease in electrical conductivity after 4000-hour exposure of a Ni-YSZ anode to a H$_2$ environment at 1000$^\circ$C. They also found a $26\%$ average Ni particle-size increase via analysis of micrographs of cross-sections. Thyden et al.~\cite{Thyden08,Hagen1} performed an aging experiment of a SOFC over $17,500$ hours. The cell was operated at 850$^\circ$C with an initial current density of 1 A/cm$^2$. Optical microscopy, field emission-scanning electron microscopy (FE-SEM), SEM-charge contrast (SEM-CC), focused ion beam (FIB)-SEM, and EIS measurements were utilized to analyze the microstructural evolution in the Ni-YSZ anode. Their results suggested that an increase in the H$_2$O concentration can promote Ni particle coarsening and lead to conductivity loss within the Ni-YSZ cermet. Tanasini et al.~\cite{Tanasini09} also conducted coarsening experiments for a single SOFC operated at 850$^\circ$C with humidified H$_2$. Although cell degradation is often attributed to cathodic processes (e.g., Cr poisoning and cathode-electrolyte interface formation reaction), they reported that the major cell performance reduction stems from the anode degradation due to coarsening.  They also found that the cell potential drop and Ni particle size increase reach a plateau after $\sim 1000$ hours of operation. Despite these experimental efforts, a quantitative correlation between microstructure and coarsening still awaits explication.

Even though the performance of SOFCs can be largely affected by microstructural changes, only a few models have been proposed to study the effects of coarsening on the performance of electrodes~\cite{Ioselevich,IoselLehnert,Vaben}. Due to the inability of acquiring 3D microstructural information for Ni-YSZ anodes, these models were mostly based on empirically fitted parameters or simplified microstructures.  The use of oversimplified microstructures and empirical parameters without extensive validations can be problematic because it can lead to incorrect conclusions.  

Recently, we have demonstrated that the microstructure of SOFC electrodes can be three-dimensionally reconstructed using dual-beam FIB-SEM~\cite{James}.  The data can be acquired to produce a wealth of information, including the tortuosity of individual phases and the TPB density.  While FIB-SEM allows the reconstruction of anode microstructure in three dimensions, this technique damages the materials. Thus, no evolution information can be obtained for the specimen after the procedure.  Modeling offers the advantage of allowing systematic analyses of coarsening effects on the performance of SOFC anodes.  

The phenomenon of anode coarsening can be described as a free-boundary problem in the sharp-interface modeling framework; however, explicit tracking of the evolving phase boundaries is highly impracticable in three dimensions.  For a 3D anode with a complex microstructure, it is therefore advantageous to use a diffuse interface approach, such as a phase-field modeling, to model the microstructural evolution. Therefore, we have developed a diffuse-interface modeling framework, based on the phase-field model and the smoothed boundary method (SBM), in conjunction with FIB-SEM experiments to quantitatively investigate Ni coarsening.

\subsection{The phase field model}
\label{PFM}

The fundamental basis of a phase-field approach is to define field variables, or order parameters (OPs), that distinguish the phases in a multi-phase system.  OPs have a constant value in each bulk phase, while interfaces are represented by finite regions in which the OPs smoothly vary from one bulk value to another. Since the boundary information is embedded in OPs, explicit tracking of moving boundaries is no longer necessary. This leads to a computational advantage in modeling multi-phase systems with multiple interfaces, especially in three dimensions.  The phase-field model, one of the phase-field approachs, can also be considered as a type of diffuse interface model, which describes interfaces using a finite thickness.

The inclusion of OPs into the free energy density is generally attributed to Landau and Ginzburg for their work related to the modeling of superconductivity in the 1950s~\cite{Landau}. Later, in order to describe the interfacial energy of an inhomogeneous system, Cahn and Hilliard~\cite{CahnHilliard} proposed to describe the free energy of a system by its OPs and their spatial derivatives.  The concept of describing the evolution of an interface between two phases differing in composition with a Ginzburg-Landau-type functional was introduced by Langer~\cite{JSLanger}. Allen and Cahn~\cite{AllenCahn} developed the theory for the motion of a coherent anti-phase boundary.  Formal asymptotic analyses have been used to show that a variety of phase-field models (PFMs) recover the corresponding sharp interface models when the width of the interfaces approaches zero~\cite{WBM,GCaginalp}.

PFMs have been successfully utilized to describe phase transition in two-phase systems (see, e.g.,~\cite{WBM}), dendrite growth (see, e.g.,~\cite{Karma97}), and Ostwald ripening (see, e.g.,~\cite{Warren96}).  However, in three-phase systems, a PFM with a generic free energy functional may introduce an artificial third-phase contribution at a two-phase boundary, if no special treatment is applied~\cite{Nestler,Folch}.  This effect can compromise the study of cermet anode coarsening since the emergence of a phantom phase at a two-phase boundary can contribute to extra TPBs.  This foreign-phase creation will lead to an erroneous estimation of the electrochemically active sites.  Nestler {\itshape et al.}~\cite{Nestler} proposed two types of remedies for this problem in a multi-component liquid-solid system.  In each of these approaches, the potentials penalize equal contributions from each of the OPs to reduce the third phase appearance.  However, the modified potentials may not eliminate the PFM simulation artifact for a simple three-phase conserved-field system undergoing coarsening.  To address this issue, Folch {\itshape et al.}~\cite{Folch} developed a specific minimal model for a three-phase system that ensures that there is no third phase invasion at a pure two-phase boundary.  This model can also accommodate systems with unequal surface tensions by adding tunable saddle-lifting terms. However, the computations become too expensive due to the steeper free-energy landscape needed for large-scale simulations. 

The PFM handles the interface implicitly because the interface information is embedded in OPs.  Immiscibility and interfacial energy are naturally incorporated in the PFM via the bulk free energy and the gradient energy penalty over the diffuse-interfacial region.  The interfacial energy ratios among different interfaces can be specified in the PFM using proper parameterization of the free energy functional. Therefore, we developed a first phase-field model, named Model A, within the context of a multiphase PFM, to model the Ni coarsening in three-phase anodes~\cite{hsunyi}. As discussed later, Model A allows a small mobility of YSZ, which significantly affects the evolution. Therefore, we developed an alternative model based on the smoothed boundary method (SBM), which is briefly described in the following section.

\subsection{The smoothed boundary method}
\label{SBM}

Complex geometries are abundant in naturally and man-made objects.  To study physical processes or phenomena occurring within such objects, the numerical solution of partial differential equations (PDEs) with prescribed boundary conditions is necessary. A standard scheme requires triangulation of the complex shapes, followed by solving of the PDEs using the finite element method (FEM). However, automatic generation of proper meshes for complex 3D domains is challenging.  In addition, in many cases the complex geometries can evolve during a physical process, thus demanding a dynamic remeshing of the evolved domain.

To address these difficulties, one can alternatively embed complex geometries within a larger, simpler domain (such as a cube). The PDEs can then be solved with regularly shaped meshes on the extended domain, provided that the original boundary conditions can be properly applied. Methods employing such a concept include composite FEM~\cite{Hackbusch97}, extended FEM~\cite{Duarte2000}, the immersed interface method~\cite{Gong08}, the immersed boundary method~\cite{Mittal05}, and the cut-cell method~\cite{Ji06}.

The smoothed boundary method differs from the above mentioned methods in that it represents the complex geometries with a phase-field-like function, which veries smoothly across the domain boundary.  Thus, the sharp boundaries of the complex geometry $\partial \Omega$ are instead described by a thin interface of a finite width. This diffuse-domain approach was utilized to study the diffusion of chemoattractant inside a cell with a no-flux boundary condition (BC) at the cell surface~\cite{Kockelkoren03}. Spectral methods were later coupled with the SBM to model electrical wave propagation in cardiac tissues with a no-flux BC~\cite{BuenoOrovio05,BuenoOrovio06}. In Ref.~\cite{hcy09}, the SBM was extended to solve PDEs in complex geometries with Dirichlet, Neumann, and Robin BCs.  An alternative but similar approach was independently developed,~\cite{XLi09}. 

In the SBM, an auxiliary variable, a domain parameter $\psi$, is introduced to identify the domain of interest $\Omega$ in which the PDEs are solved. The domain parameter $\psi$ usually has a value of $1$ within $\Omega$, $0$ exterior to $\Omega$, and $0< \psi <1$ in $\partial \Omega$.  Since the complex geometry is embedded in a regularly shaped, expanded domain, the exterior of the domain of interest is included only to facilitate computation; any numerical solutions obtained in the external region are devoid of meaningful information.  The initial construction of the domain parameter $\psi$ can be achieved by solving a phase-field equation or by transforming the distance function of the domain structure with the hyperbolic tangent.

It is generally believed that the YSZ phase has a very low mobility in an operating SOFC anode; that is, YSZ serves as the supporting structure in which Ni coarsens. We therefore propose a model, named Model B, utilizing the SBM for Ni coarsening simulations. In Model B, we assume that YSZ is stationary. It is therefore treated as the geometry within which the Ni and pore phases evolve.  We also assume that the triple junctions in the Ni-YSZ anode possess the contact angles deduced from Young's equation for a locally flat surface.  The dynamics of this system can thus be modeled by a single OP Cahn-Hilliard equation with two complementary BCs implemented with the SBM: the contact-angle BC at triple junctions and the no-flux BC at YSZ interfaces.

\section{Methods}
In this section, we elaborate the model framework that we previously published in a short communication for Ni coarsening~\cite{hsunyi}. The non-dimensionalization, asymptotic analysis, and numerical methods are discussed in detail.  Moreover, the methodology we used to identify the TPBs and distinguish the percolated or isolated phases is discussed.  An error analysis is performed to examine how the ratios between the characteristic length and the interfacial thickness affect the contact angles at triple junctions in our model. 

\subsection{Model Formulation}
We have proposed two PFMs for coarsening simulations in three-phase SOFC cermet anodes.  Our models are based on a free energy funtional that is computationally inexpensive and can circumvent the phantom phase issue associated with some other solutions.  The approximations we made in both of our models are: (1) the volume (mass) of each phase in the Ni-YSZ anode is conserved so that the Cahn-Hilliard dynamics is applicable; (2) material properties required for modeling are estimated at $1000^\circ$C; (3) Ni surface free energy is assumed to be isotropic; (4)  Ni surface diffusivity is estimated by the mean value of multiple crystal orientations; (5) surface diffusion dominates. 

In our PFMs, the Cahn-Hilliard evolution equations can be considered as a type of diffusion equation, where the material transport is driven by the chemical potential. This potential can be formulated by the design of a free energy functional that is based on the effective driving forces for material transport.  At ordinary operating temperatures, the three phases comprise the Ni-YSZ anode can be assumed immiscible; therefore, lattice misfit as well as elastic energy contribution toward the diffusion are negligible.  In addition, external forces such as gravity are not considered. We thus formulate our free energy functional as the simplest Ginzburg-Landau type functional $ F = \int\limits_{V} f dV$, where the free-energy density $f$ simply depends on the OPs and their derivatives. 

The two PFMs we have proposed differ in their treatment of the YSZ phase. If we consider that the YSZ phase can transport with its mobility, the Ni-YSZ anode is then a system of three mobile phases (Ni, YSZ and Pore). This system requires two evolution equations to resolve the kinetics. We call this model {\itshape Model A}.  In {\itshape Model B}, we consider the YSZ phase to be completely immobile, which is justified because its mobility is orders of magnitude smaller than that of the Ni phase.  Therefore, only two phases, Ni and pore, are allowed to evolve in between of the YSZ matrix.  To incorporate the YSZ phase as the internal boundary in the computational domain, we have developed a model utilizing the smoothed boundary method. 

\subsubsection{Model A}
In Model A, the OPs are vectorized as $\overrightarrow{\phi} = (\phi_{1},\phi_{2},\phi_{3})$ for generalization. Each vector component represents the volume fraction of the corresponding phase so that an additional constraint $\phi_{1}+\phi_{2}+\phi_{3}=1$ applies; thus, only two evolution equations are needed.  The governing equation sets can be written as
\begin{equation}
\begin{array}{l}
\displaystyle	\frac{\partial \phi_{1}}{\partial t}=\triangledown \cdot M(\overrightarrow{\phi}) \triangledown \frac{\delta F}{\delta \phi_{1}},\\
\displaystyle \\
\displaystyle	\frac{\partial \phi_{2}}{\partial t}=\triangledown \cdot M(\overrightarrow{\phi}) \triangledown \frac{\delta F}{\delta \phi_{2}},
\end{array}
\label{eq:eqCH2}
\end{equation}
where $M(\overrightarrow{\phi})$ is the surface mobility.

A standard bulk free energy for a three-phase system should contain three local minima, each of which represents a bulk-phase value.  We select the three local minima as $\Phi_{a} = (1,0,0), \Phi_{b} = (0,1,0), \Phi_{c} = (0,0,1)$. However, in the three-phase Cahn-Hilliard dynamics, a foreign phase can be introduced at a two-phase interface. Two remedies are that one can use the procedure developed in~\cite{Folch} to generate a specific free energy, or one can use an interpolation function similar to that described in~\cite{Moelans2010} (the latter method may only reduce the amount of the third phase contribution). Since the three phases in the Ni-YSZ anodes are immiscible, the generation of a foreign phase at a two-phase boundary is a problematic artifact that leads to extra TPB sites that do not physically exist.  

Our approach to resolve this issue is to select a free energy functional that can limit a foreign-phase appearance and can also incorporate an unequal surface tension at a reasonable computational expense. In the Cahn-Hilliard dynamics, the excess free energy at the interfaces, i.e., the interfacial energy, is determined by gradient energy coefficients and the bulk free energy at intermediate values of OPs. According to Young's relation, the contact angles among the different phases are determined by the interfacial energies of the intersecting interfaces.  For instance, if phases 1, 2, and 3 are in contact at a triple junction, the contact angle formed in phase 1 is related to the interfacial energy of the interface 2-3 relative to that of the interfaces 1-3 and 1-2.  It is, however, easier to understand the interfacial energy as a factor reflecting the affinity of one phase to another. A lower interfacial energy suggests a stronger bonding of the two phases that are in contact. This tendency appears to have profound effects on the long-term microstructural evolution as demonstrated in our simulations~\cite{hsunyi}.

Applying the commonly used gradient and bulk free energy densities for multiple-OP systems, our free energy functional can be written as

\begin{equation}
	F =  \int \limits_V dV \left ( \sum_{i=1}^{3} k_{i} \phi_{i}^{2} (1-\phi_{i})^{2} -\sum_{i,j=1,i>j}^{3} \alpha_{ij}^{2}\nabla \phi_{i} \nabla \phi_{j} \right ) .
	\label{eq:FEF}
\end{equation}

The specific interfacial energy is calculated from the equilibrium solution at a planar interface.  To do so, the free energy functional must be minimized subject to a constraint that the sum of the order parameters equals unity.  This is carried out by taking the variational derivative of the functional using the method of Lagrange multipliers.  The resulting equation is given by 

\begin{equation}
	\left ( \frac{\delta F}{\delta \phi_{i}} \right )_{\phi_{1}+\phi_{2}+\phi_{3}=1}= \frac{\delta F}{\delta \phi_{i}} - \left( \frac{1}{3} \right ) \sum_j \frac{\delta F}{\delta \phi_{j}},
	\label{eq:DfDphi}
\end{equation}
 where the variational derivatives at the right hand side of the equation are taken as if all $\phi_i$'s were independent~\cite{Folch}.
If we consider an interface between phase $i$ and phase $j$ without the existence of the third phase, the interfacial free energy can be calculated from the integration of the free energy density over the interface as
\begin{equation}
     \gamma_{ij} = 2 \alpha_{ij} \sqrt{(k_{i}+k_{j})} \int_{0}^{1} \phi_{i} (1-\phi_{i}) d\phi_{i} = \frac{\alpha_{ij}}{3} \sqrt{(k_{i}+k_{j})}.
 \label{eq:SurfE}
\end{equation}
Similarly, the interfacial width $\delta_{ij}$ can also be found as
\begin{equation}
     \delta_{ij} = \frac{\alpha_{ij}}{ \sqrt{(k_{i}+k_{j})}}.
 \label{eq:SurfW}
\end{equation}
After multiplying Eq.~\ref{eq:SurfE} by Eq.~\ref{eq:SurfW} and rearranging the result, we find 
\begin{equation}
 3 \delta_{ij} \gamma_{ij} = \alpha_{ij}^2. 
\label{eq:SurfEW}
\end{equation}
By choosing $\delta_{12}=\delta_{23}=\delta_{13}$, the relationships between the gradient energy coefficients and the interfacial energy in this three-phase system is simplified as

\begin{equation}
	\frac{\gamma_{12}}{\alpha_{12}^{2}} = \frac{\gamma_{23}}{\alpha_{23}^{2}} = \frac{\gamma_{13}}{\alpha_{13}^{2}}.
    \label{eq:SFEGEC}
\end{equation}
The values of the surface free energies can be set to those found in the literature or be left as a parameter if data are missing or uncertain.  The interfacial width is a computational parameter that must be chosen to ensure thin-interface limit for accuracy while providing a numerically well-resolved interface. Once the interfacial width $\delta_{ij}$ and interfacial energies $\gamma_{ij}$ are set, the gradient energy coefficients $\alpha_{ij}$ can be calculated from Eq.~\ref{eq:SurfEW}, and the bulk energy coefficients $k_i$ can be calculated from Eq.~\ref{eq:SurfW}.

Surface diffusion is generally considered to be the main mechanism of transport for the capillarity-driven microstructural evolution in these SOFC anodes~\cite{Vaben}. We have confirmed the validity of this assumption based on an analysis of order of magnitudes.  Since the OPs with integer values represent the bulk phases in the PFMs, the surface mobility function in a one-OP, two-phase case is commonly formulated as $M(\phi) = \phi^2 (1-\phi)^2$; such a formulation guarantees a non-zero mobility at two-phase boundaries only.

In Model A, the surface mobility function is given by:
\begin{equation}
M(\phi_{1},\phi_{2},\phi_{3}) =\sum_{i,j=1,i>j}^{3} M_{ij} \textstyle \prod_{C_{l}C_{h}}(\phi_{i}) \prod_{C_{l}C_{h}}(\phi_{j}) \big ( \phi_{i} \phi_{j} (1-\phi_{i})(1-\phi_{j}) \big ), \label{eq:mobF}
\end{equation}
 where we introduce a boxcar function $\prod_{C_{l}C_{h}}(\phi_i)$ to avoid excess mobility resulting from the appearance of a small amount (less than 3\% in fraction) of a foreign phase at a two-phase boundary.  We choose $C_{l}=0.05$ as the lower cutoff OP value in the mobility function, and $C_{h}=1-C_{l}$ as the upper cutoff value and resulting from the complementary value of the lower cutoff.  All three mobility prefactors have positive values, and thus, materials flow from high to low chemical potentials.

\subsubsection{Model B}
In this model, we assume that YSZ is stationary and that the Ni phase evolves by diffusion along the Ni-pore interfaces.  The dynamics of this system can thus be described by a single-OP Cahn-Hilliard equation with one contact-angle BC at triple junctions, and one no-flux BC at YSZ interfaces, where the OP distinguishes the Ni and pore phases.

The treatment of a contact-angle BC in a non-conserved (Allen-Cahn) PFM was proposed by Warren {\itshape et al.} \cite{Warren} to model heterogeneous nucleation.  Following \cite{Warren}, we developed Model B using the SBM for a Ni-YSZ anode coarsening with conserved dynamics.  This SBM approach is an entirely diffuse-interface treatment that implicitly handles complex geometries (while Warren et al. applied delta function~\cite{Warren}).  In the following derivation, we demonstrate that both no-flux and contact-angle BCs can be coupled with a Cahn-Hilliard equation to give a single evolution equation using the SBM framework.

For one-OP Cahn-Hilliard dynamics within a SBM framework, the evolution equation can be written as
\begin{equation}
     \frac{\partial \phi}{\partial t} = \nabla \cdot M(\phi,\psi) \nabla \mu.
	\label{eq:CHev}
\end{equation}
We consider the free energy functional as being:
\begin{equation}
    \mathcal{F} = \int_V dV [\frac{\epsilon^{2}}{2} \left | \nabla \phi \right \vert^{2} + f(\phi)],	
	\label{eq:Teng}
\end{equation}
where $f(\phi)$ is a generic double-well function.
The chemical potential $\mu$ is by definition the variational derivative of free energy $\mathcal{F}$ with respect to the order parameter $\phi$, {\itshape i.e.}, $\mu = \delta \mathcal{F} /\delta \phi = \partial f /\partial \phi - \epsilon^2 \nabla^{2} \phi$. 
We introduce in the SBM a domain parameter $\psi$ to incorporate general BCs on internal boundaries.  In this case, the domain parameter $\psi$ distinguishes regions having YSZ ($\psi=0$) and other phases ($\psi=1$), as well as the YSZ interfaces ($0<\psi<1$).

Multiplying Eq.~\eqref{eq:CHev} by $\psi$ and letting $\vec J = M \nabla \mu$ the formulation becomes
\begin{equation}
    \frac{\partial ( \psi \phi)}{\partial t} = \psi \nabla \cdot \vec J = \nabla \cdot (\psi \vec J) - \vec J \cdot \nabla \psi.
    \label{eq:CHevM}
\end{equation}
The aforementioned no-flux BC, $\nabla \mu \cdot \nabla \psi = 0$, should be applied to the internal boundaries (YSZ interfaces) to ensure mass conservation.  This no-flux BC eliminates the last term of Eq.~\eqref{eq:CHevM} and the equation becomes
\begin{equation}
    \frac{\partial ( \psi \phi)}{\partial t} = \nabla \cdot (\psi \vec J) = \nabla \cdot (\psi M \nabla \mu).
    \label{eq:CHm1}
\end{equation}

Like an ordinary OP in a phase-field approach, $\psi$ varies continuously across the interface; thus, the unit interface normal $\vec n$ of the YSZ interface can be described as a function of the gradient of $\psi$, i.e., $\vec {n} = \nabla \psi / \left | \nabla \psi \right \vert$.  By assuming that the effects of the YSZ on the Ni phase are only of short range (immiscible) and $\phi=1$ represents the bulk Ni phase, the contact angle $\theta$ at triple junctions can be formulated as 
\begin{equation}
    \vec n \cdot \frac{\nabla \phi}{\left | \nabla \phi \right \vert} = \frac{\nabla \psi}{\left | \nabla \psi \right \vert} \cdot \frac{\nabla \phi}{\left | \nabla \phi \right \vert} = -\cos \theta.
    \label{eq:AngBC}
\end{equation}
The negative sign comes from the convention that $\nabla \psi$ points into YSZ and $\nabla \phi$ points out of Ni.

The mechanical equilibrium at the triple junction corresponds to an extremum of the free energy, i.e., $\delta \mathcal{F} = 0$.  We can use the planar solution of the thermodynamic equilibrium condition within the interface to find a useful equality, $\left | \nabla \phi \right \vert = \sqrt{2 f} / \epsilon$, which can be substituted into Eq.~\eqref{eq:AngBC} to derive the SBM contact-angle BC as
\begin{equation}
    \nabla \psi \cdot \nabla \phi = - \left | \nabla \psi \right \vert \cos \theta \frac{\sqrt{2f}}{\epsilon}.
    \label{eq:AngBCF}
\end{equation}
This contact-angle BC results in energy only near triple junctions, rather than in the bulk volume, to achieve the force balance at the triple junction that is dictated by Young's equation for a flat surface. The final evolution equation derived from the SBM with no-flux and contact-angle BCs is: 

\begin{equation}
\frac{\partial ( \psi \phi)}{\partial t} = \nabla \cdot \bigg \{ \psi M \nabla \bigg [ f_{\phi} - \frac{\epsilon^{2}}{\psi} \bigg ( \nabla \cdot ( \psi \nabla \phi ) + \frac{\left | \nabla \psi \right \vert  \sqrt{2f}}{\epsilon} \cos \theta   \bigg ) \bigg ] \bigg \}.
\label{eq:CHSBMF}
\end{equation}
Again, the gradient energy coefficient $\epsilon$, the contact angle $\theta$, and the bulk energy coefficients in $f(\phi)$ are determined according to the interfacial energies of the Ni-YSZ cermet and the selected interfacial widths.

The mobility function in Model B is formulated as 
\begin{equation}
M(\phi,\psi) = M_{Ni-Pore} \textstyle \prod_{C_{l}C_{h}}(\phi) \big ( \phi^2 (1-\phi^2) \big ) g(\psi),
\label{eq:SBMMob} 
\end{equation}
 where $g(\psi) = \psi^6 (10\psi^2-15\psi+6)$ is introduced to control the mobility at and near triple junctions.  This one-sided interpolation function $g(\psi)$ transitions smoothly from 1 to the order of $0.01$ as the domain parameter varies from 1 to $0.5$. In other words, this choice of the mobility function ensures an immobile YSZ phase by limiting the mobilities at YSZ interfaces. Subsequently, the mobility near a triple junction decreases from the Ni-pore value to a value that is $~10^{-6}$ smaller as $\psi$ varies from 1 to about $0.1$.

\subsection{Nondimensionalization and Asymptotic Analysis}
Having the appropriate values of the mobility prefactors is essential in simulating coarsening kinetics. In order to quantitatively correlate simulation results with physical phenomena, asymptotic analyses are required in PFMs because of their diffuse interface nature. Using an asymptotic analysis, we determined the relationship between diffusivity and mobility and the characteristic simulation time scale. Unlike some of the previous studies, which determined model parameters by fitting coarsening experimental results, our model is a predictive model that is free of fitting parameters when the material properties, such as surface diffusivities, are accurately known.  

In the Ni-YSZ anode of SOFCs, coarsening proceeds mostly via surface diffusion. The anisotropic effect of crystal facets on surface diffusion is lumped into one ensemble diffusivity in our models.  In this case, the normal velocity $V_{n}$ of the interface $\Gamma$ incurred by surface diffusion, in the dimensional sharp-interface form, can be represented as
\begin{equation}
	V_{n} = \frac{\gamma_{s} D_{s} \delta_{s}}{k_{B} T N_{v}} \nabla_{s}^2 \kappa_{c} = \frac{\gamma_{s} D_{s} \delta_{s}}{k_{B} T N_{v}} \frac{\partial^{2} \kappa_{c}}{\partial s^{2}},
	\label{eq:SHP}
\end{equation}
, where $\gamma_{s}$ is the surface energy, $D_{s}$ is the surface diffusion coefficient, $\delta_{s}$ is the interfacial thickness, $N_{v}$ is the atomic number density per volume, $\kappa_{c}$ is the local curvature, $\nabla_{s}^2$ is the surface Laplacian and $\partial / \partial s$ is the gradient operator along the interface~\cite{Balluffi2005}.

To link the corresponding diffuse-interface model to the aforementioned sharp-interface model, a pure two-phase boundary in this three-phase system is considered. For both Model A and Model B, the dimensional Cahn-Hilliard evolution equation at a two-phase boundary can be re-arranged as
\begin{equation}
	\frac{\partial \phi}{\partial t} = \triangledown \cdot ( M(\phi) \triangledown \mu ),
	\label{eqDCH}
\end{equation}
\begin{equation}
	\mu = \frac{\partial f(\phi)}{\partial \phi} - \epsilon^{2} \triangledown^{2} \phi,
	\label{eqDmu}
\end{equation}
, where $\phi$ is the OP distinguishing two phases, $f(\phi) = W \phi^{2}(1-\phi)^{2}/4$ is the energy density, and $M(\phi) = 6M_{s} \phi^2(1-\phi)^2$ is the surface mobility.  Eq.~\ref{eqDCH} and Eq.~\ref{eqDmu} possess the physical steady-state solution for a planar interface, i.e., $\phi(x)=[1-\tanh(x/2 \delta)]/2$, with the interface thickness $\delta = \epsilon \sqrt{(2/W)}$, and the interfacial energy $\gamma = \epsilon \sqrt{(W/72)}$. 

Following the derivation in~\cite{SteveWise}, a new set of variables for a non-dimensionalization of the governing equations is utilized. We assume that $\tau = L^{4}/D$, wherein $L$ is the characteristic length scale of the sample and $\tau$ is the characteristic time scale for surface diffusion, and consider that other nondimensional quantities (denoted by overbars) are defined by
\begin{eqnarray}
&&	\bar{M}_{s} = \frac{36 M_{s} \gamma \delta}{D},~~ \bar{\delta} = \frac{\delta}{L},~~ \bar x = \frac{x}{L},~~\bar t = \frac{t}{\tau} \nonumber \\
&& \bar{M}(\phi) = \phi^{2}(1-\phi)^{2},~~ \bar{f}(\phi) = \frac{1}{2} \phi^{2}(1-\phi)^{2}.
\label{eq:NDParm}
\end{eqnarray}
Eqs.~\ref{eqDCH} and~\ref{eqDmu} can be written using the non-dimensional parameters to derive the nondimensional evolution equation as
\begin{equation}
    \frac{\partial \phi}{\partial \bar t} =\frac{1}{\bar \delta^{2}} \triangledown \bar M(\phi) \triangledown \bar \mu
	\label{eq:NDCH3}
\end{equation}
\begin{equation}
	\bar \mu = \frac{\partial \bar f(\phi)}{\partial \phi} - \bar \delta^{2} \triangledown^{2}\phi,
	\label{eq:NDmu}
\end{equation}

The asymptotic analysis derivation is similar to~\cite{DongHee} and is detailed in the appendix. Comparing the dimension-restored equation derived from the asymptotic analysis with the corresponding sharp interface equation, Eq.~\ref{eq:SHP}, we find
\begin{equation}
	\frac{D \bar{M_{s}}}{36} = M_{s} \delta \gamma = \frac{\gamma_{s} D_{s} \delta_{s}}{k_{B} T N_{v}},
	\label{eq:MS}
\end{equation}
, which indicates that $M_{s}=D_{s} \delta_{s}/k_{B}TN_{v} \delta$, provided the choice of $\gamma = \gamma_{s}$; that is, we find that the surface mobility is proportional to surface diffusivity. Choosing $\bar{M}_{s} = 36$, the model variable $D$ is connected to the surface diffusivity $D_{s}$ as follows:
\begin{equation}
	D = \frac{\gamma_{s} D_{s} \delta_{s}}{k_{B} T N_{v}}.
	\label{eq:DDs}
\end{equation}
The time scale that links the simulation time to the physical time is acquired from Eq.~\ref{eq:DDs} as
\begin{equation}
	\tau = \frac{L^{4} k_{B} T N_{v}}{\gamma_{s} D_{s} \delta_{s}}.
	\label{eq:Tau2}
\end{equation}

\subsection{Numerical methods}
One of the challenges in modeling Ni-YSZ anode coarsening is the fact that solving Cahn-Hilliard equations with an explicit time iteration scheme is too expensive, especially for a large scale simulations in three dimensions.  Thus, we solve the nondimensional evolution equations, Eqs.~\ref{eq:NDCH3} and~\ref{eq:NDmu}, with the algorithm based on splitting the fourth-order Cahn-Hilliard equation into two second-order equations and soloving for the OPs and chemical potential simultaneously~\cite{SteveWise}.  We use the central-differencing method for the spatial discretization and the Crank-Nicholson algorithm for the time discretization. This semi-implicit scheme significantly reduces the stiffness of the numerical integration, which allows much larger time step size.    

To solve the nonlinear finite-difference equations, Newton's method is utilized for the nonlinear terms. A pointwise Gauss-Sidel relaxation scheme and a red-black checkerboard iteration scheme are used together to accerelate the convergence rate as well to facilitate the parallelization.  In most cases, we find this solver is over 100 times faster in comparison to the explicit scheme.  However, attention must be paid to the choice of the time stepping size.  We find in some cases, when an overly large time step is used in our solver, the simulation results are incorrect even if the numerical scheme is stable.

When solving the coupled governing equations in Model A, the solver is parallelized with Message Passing Interface (MPI) library to take advantage of the multiple processors.  In our MPI code, the domain is decomposed equally in size in each axial direction, if possible, to achieve load balance.  For instance, if 64 CPUs are allocated to the computation, the entire domain is decomposed into 4 by 4 by 4 sub-domains; that is, each sub-domain has a domain size of 1/64 of the original domain in volume or, more explicitly, 1/4 of the length of the original domain in each axis (assuming there are no residuals). 

The solver for Model B is parallelized with both the MPI and openMP libraries. Since only one Cahn-Hilliard equation is solved, it is much more numerically stable and efficient compare to the Model A solver.

\subsection{Error analysis: contact angles and the interfacial width}
The phase-field model is known to smooth out microstructures with length scales below the diffuse interface thickness. This artificial smoothing process introduces errors during the early stages of a simulation, but has no negative effects on the analysis of long-term coarsening kinetics. In contrast, microstructural evolution kinetics can only be accurately resolved in PFMs when the length scale of the microstructure is larger than certain multiples of the interfacial thickness. In other words, there is a critical ratio between the microstructural length scale and the interfacial thickness that is required to obtain a sufficient agreement with the sharp-interface limit. For example, if a system contains particles with typical radii a few times smaller than the interface thickness, the simulations of its evolution will incur a large error.  The commonly recognized critical ratio is of about 10. However, in large scale 3D simulations, to resolve a complicated system with microstructural features of different length scales, achieving this critical value for all features is impracticable because it would require a very high resolution.  In practice, we use a smaller ratio, especially when there are smaller features among a range of feature sizes within the microstructure, and quantify the errors introduced by the selected value. 

Because the coarsening kinetics of Ni-YSZ anodes has been found to depend strongly upon the Ni-YSZ contact angle, an error analysis is performed via the investigation of the contact angle of Ni on the YSZ phase in two dimensions (2D).  The two-dimensional domain is initialized with a bottom that is half occupied by YSZ and a top half that is equally divided into the Ni and pore phases with a $90^\circ$ contact angle. The remainder of the domain is filled with the pore phase.  The domain size is designed to be large enough so that no-flux BCs have negligible effects on the contact angles at the triple junction. The system is evolved with our Model B to its steady state.  The final contact angle $\theta_C$ is calculated from the average value of the dot product of the normal,
\begin{equation}
    \frac{\nabla \psi}{\left | \nabla \psi \right \vert} \cdot \frac{\nabla \phi}{\left | \nabla \phi \right \vert} = -\cos \theta_C,
    \label{eq:ctAng}
\end{equation}
over the region where the domain parameter $\psi$ and the order parameter $\phi$ are both between 0.1 and 0.9, indicating the TPB region. 

Two contact angles are studied: $120^\circ$ and $93^\circ$.  The $120^\circ$ case is selected as a reference case because the cosine function is away from the extrema or inflection point at this contact angle.  The $93^\circ$ case corresponds to a physical contact angle of Ni on the YSZ phase that is based on our selected interfacial energies.  

\begin{figure}[htb!]
   \centerline{\includegraphics[width=10.37cm,height=7.5cm]{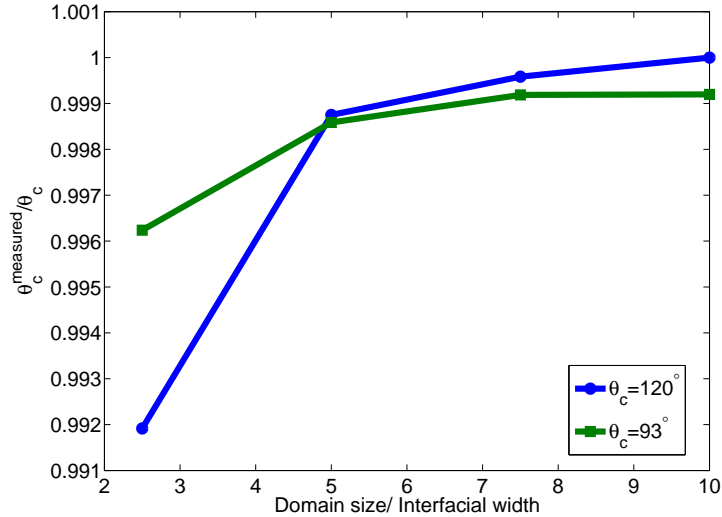}}
	\caption{Normalized angle plotted with the ratio of the domain size to the interfacial width. The error decreases with an increasing ratio of the domain size to the interfacial width. \label{fig:ErrAng}}
\end{figure}

As the characteristic length of the system, we choose the domain size, which is varied from 10 points to 100 points in each direction, while the interfacial width is held at 4 grid points.  Therefore, the ratio of the domain size to the interfacial width varies from 2.5 to 10.  For the $10\times 10$ domain, the interfacial region (either the domain or order parameters are between 0.1 and 0.9) occupies about half of the domain. As the domain size increases, the fraction becomes very small.  The contact angles normalized to the set value versus the ratios of the domain size to the interfacial width are plotted in Fig.~\ref{fig:ErrAng}. In the case of a $120^\circ$ contact angle, we find that the angle deviate from the sharp-interface value by less than $0.2^\circ$ when the ratio is larger than $5$.  Even at a ratio of $2.5$, the deviation is below $1$ degree or $1\%$.  In the case of a $93^\circ$ contact angle, the contact angle differs from the sharp-interface value by less than $0.5^\circ$, or $0.5 \%$.  

The interfacial width utilized in coarsening simulations is approximately $0.13~\mu$m.  Based on our error analysis results, the contact angles of the Ni particles in the active and dead-end categories with a length scale larger than $0.32~\mu$m can fairly accurately be modeled by using our Model B. In the isolated network, a length scale larger than $0.64~\mu$m is needed for a fair accuracy.  For the $(4~\mu$m$)^3$ domain, only $3\%$ in volume of the Ni phase belongs to clusters outside the accurate range.  In addition, based on Young's equation for a flat surface, a $0.5^\circ$ difference in contact angle of the Ni phase corresponds to a $1.1\%$ difference in Ni-YSZ interfacial energy. In other words, this model error has very limited effects on the coarsening kinetics of the Ni-YSZ anode. 

\section{Parameterization of the models}
\label{ModelPara}

The aforementioned asymptotic analysis provides the mathematical ground for correlating simulations on the coarsening phenomenon with experimental results. However, to quantitatively model the physical process, the material-specific parameters in the governing equations must be specified based on the material properties. Most of these parameters can be found in the literature. 

\subsection{Mobility prefactors}
As demonstrated in our asymptotic analysis, the mobility prefactor corresponding to the Ni-pore interface should be proportional to the surface self-diffusivity of Ni, which is usually anisotropic or depends on the crystallographic orientation.  The Ni diffusivities based on the field ion microscope (FIM) measurements range from $10^{-13}~$m$^{2}$s$^{-1}$ in (110)~\ to $10^{-9}~$m$^{2}$s$^{-1}$ in (331) at 1273 K~\cite{Seebauer,Tung,Fu}; these values are extrapolated at high temperatures and could be inaccurate, because FIM can be conducted only at low temperature regions ($T < 0.2 T_{m}$, $T_m$ is the melting temperature).  On the other hand, the surface smoothing method (SSm) measures mass-transfer diffusion at high temperatures ($T > 0.7 T_{m}$), which is generally averaged over crystallographic orientations~\cite{Seebauer}.  The latter method provides a more appropriate value for the diffusivity ($\sim 10^{-10}~$m$^{2}$s$^{-1}$) during Ni coarsening at the modeled temperature (1000$^\circ$C).  In addition, the anisotropy of the Ni surface diffusivity was found to be relatively small in high temperature regions ($0.82~T_{m}< T < T_{m}$) in~\cite{Maiya}.  We thus consider an ensemble value $\sim 10^{-10}~$m$^{2}$s$^{-1}$ for the Ni surface diffusivity in our model.

Cation diffusion in oxides is related to various parameters, such as valence, atomic radius, impurity, and oxygen activity. In Ref.~\cite{Kilo}, it is reported that the bulk diffusivity of yttrium in YSZ (containing 10 to 32 mol\% Y$_{2}$O$_{3}$) is slightly larger than that of zirconium in YSZ, and that the difference is less than an order of magnitude.  The surface diffusion of Zr in YSZ is calculated from the measurements of the surface area reduction in powder compacts during sintering in Ref.~\cite{Mayo}, and the diffusivity at 1000 $^\circ$C is found to be $\sim 10^{-16}~$m$^{2}$s$^{-1}$. We thus consider a surface diffusivity of YSZ of $10^{-16}~$m$^{2}$s$^{-1}$ in our simulations.

The diffusion mechanisms at metal-ceramic interfaces are less understood than those on metal or ceramic surfaces.  The cohesiveness of the Ni-YSZ interface, which depends on the process used to fabricate the porous cermets, determines the interfacial structure and affects the effective interfacial diffusivity.  A series of diffusion bonding experiments indicated that the metal-ceramic interface does not act as an efficient vacancy sink or mass transport path~\cite{Derby}.  These findings suggest that the diffusivity at the Ni-YSZ interface should be much smaller than that of the Ni surface and the YSZ surface. In addition, the exact value of the diffusivity at the Ni-YSZ interface is not as important as that of Ni because the redistribution of Ni is observed to be the dominant morphological change during anode coarsening. The diffusivity of the Ni-YSZ interface is considered to be approximately $\sim10^{-20}~$m$^{2}$s$^{-1}$.  

Using these values, the mobility ratios among the three materials are set at $M_{\mathrm{NiP}}:M_{\mathrm{YP}}:M_{\mathrm{NiY}}=1:10^{-6}:10^{-10} $, where the subscript Ni represents nickel, Y the YSZ and P the pore phase. For non-dimensionalization, a mobility scale of $10^{-10}~$m$^{2}$s$^{-1}$ is used.

\subsection{Bulk and gradient energy coefficients}
In a PFM, the balance between the bulk and the gradient energy terms in the free energy functional determines the thickness of the interface.  The ratio of the interfacial thickness to the characteristic length of the system and the ratio of the domain size to the characteristic length are crucial for the validity of a PFM.  It is known that an interface that is too thin can cause an unphysical pinning or halting of coarsening, whereas an interface that is too thick can lead to unphysical dissolution of particles~\cite{NingMa2006}.  Therefore, our task is to select bulk and gradient energy coefficients that result in an appropriate interfacial energy, while keeping the interfacial thickness sufficiently large so that the computation is feasible.

In a multi-phase system, using an unequal interfacial thickness among different interfaces without a special treatment can cause erroneous evolution kinetics in a PFM simulation because the interfacial thickness changes the ratio of the simulation time scale to the physical time scale. Therefore, a single value for the interfacial thickness of all interfaces is set. For example, we set $\delta_{ij} = \Delta x$, which gives 4 points in the interfacial region to avoid the aforementioned pinning while optimizing the computational efficiency ($\Delta x$ depends on the selected length scale in a non-dimensionalization).  The interfacial energies are obtained from existing experimental or computational studies.  These choices are then used in Eqs.~\ref{eq:SurfW} and~\ref{eq:SurfEW} to determine the bulk and gradient energy coefficients.

The three interfaces that exist in the Ni-YSZ anode are the Ni surface (Ni-pore interface), the YSZ surface (YSZ-pore interface), and the Ni-YSZ interface.  The Ni surface free energy has been widely investigated and the values reported are in good agreement.  At 1000$^\circ$C, the surface energy of $\sim 1.9~$Jm$^{-2}$ for Ni has been measured~\cite{Mantzouris05}. A 8YSZ surface (8 mol \% Y$_{2}$O$_{3}$) has been reported in Ref.~\cite{Tsoga}. It was studied at 1300$\sim$1600$^\circ$C using a multiphase equilibrium technique. The YSZ surface free energy $\gamma_{\mathrm{YSZ}}$ was found to decrease linearly from 1.26 to 1.13 Jm$^{-2}$ in the range of temperatures studied. By extrapolating this data, one obtains $\gamma_{\mathrm{YP}} \sim 1.4~$Jm$^{-2}$ at 1000$^\circ$C. This value is used to parameterize our model. In Ref.~\cite{Ballabio}, ab initio calculations have been reported with $\gamma_{\mathrm{YP}}$ ranging from 1.04 to 1.75 Jm$^{-2}$ at $T = 0$, depending on the crystallographic orientation.

The free energy of a heterogeneous Ni-YSZ interface depends on the interfacial structure, which is dependant on the fabrication process. Nikolopoulos {\itshape et al.} experimentally measured the non-reactive contact angle between molten Ni and 8YSZ at 1500$^\circ$C and found a value of 117$^\circ$, which suggests an interfacial energy of 1.95 Jm$^{-2}$~\cite{Nikolopoulos96}.  Although this wetting experiment was not conducted at normal SOFC operating temperatures, the result implies a poor wettability at the Ni-YSZ interface. 

Several ab initio calculations on bond formation at Ni-YSZ interfaces have been reported~\cite{Christensen}.  A strong bonding between Ni-Zr and Ni-O was found at the Ni$(100)/$ZrO$_{2}(100)$ polar interfaces, and an interfacial tension of $\sigma_{(100)}=1.04~$Jm$^{-2}$ was reported. Another local minimum appeared at the ZrO$_{2}/$Ni$(111)$ interface where $\sigma_{(111)}=1.80~$ Jm$^{-2}$~\cite{Beltran}.  However, Ni-YSZ interfaces formed subsequent to sintering are multifaceted.  Despite the lack of understanding of these interfaces, we know the range of reasonable interfacial energies at Ni-YSZ interfaces. In our simulations, $\gamma_{\mathrm{NiY}}=1.50~$Jm$^{-2}$ is used.

In view of the above discussion, we assume that the gradient energy coefficient ratios are $\gamma_{\mathrm{NiP}} : \alpha_{\mathrm{NiY}}^2 : \alpha_{\mathrm{YP}}^2 = 1.9:1.5:1.4$ based on a surface energy scale of 1 Jm$^{-2}$, which leads to the bulk energy coefficient ratios of $k_{\mathrm{Ni}} : k_{\mathrm{Y}} : k_{\mathrm{P}} = 1:0.5:0.9$.

\section{3D data analysis}
\label{PconTPBact}

In three dimensions, a boundary at which three phases coincide has only one degree of freedom in space, i.e., the TPB is a line in the Ni-YSZ anode where the Ni, YSZ, and pore phases are all in contact.  However, to identify the TPBs based on the microstructure reconstructed by FIB-SEM, some data processing is necessary since the microstructural data are represented by a 3D matrix. In this matrix, each phase is represented by regions in which voxel values are equal to a pre-determined constant. There are several methods to determine the TPB length in 3D volumetric data. For example, in Ref.~\cite{JWilson}, the TPBs are identified as the edges where three different voxels, each of which belongs to a different phase, are in contact.  With some geometric corrections, the TPB length can be acquired with fair accuracy, but is limited by the resolution.  

In our work, we adopt a thinning, or skeletonization, algorithm to determine the TPB regions, and count the voxels. In PFM simulations, the interfacial region is commonly identified as the zone over which an OP varies from 0.1 to 0.9. Because the interfaces in PFMs span multiple grid points, one can identify the diffuse TPB regions as voxels in which all three OPs are between 0.1 and 0.9. For Model B, the threshold values are slightly different due to the additional pre- and post- treatment of the data (Ref.~\cite{HsunYi11}).  In order to recover the one-dimensional nature of TPBs from this data, we use a thinning algorithm developed in Ref.~\cite{Palagyi}, which reduces the diffuse TPB regions to the corresponding skeleton of chains of voxels. The key feature of this algorithm is that it preserves the topological characteristics of the original image/voxelated data (i.e., it does not allow the pinching or connection of regions).  

The final TPB length is calculated via a multiplication of the physical grid size and the total number of TPB voxels after skeletonization.  The used procedures may either overestimate or underestimate the TPB lengths depending on the angle at which the TPB lies within the grid; for an isotropic distribution of lines, the over/underestimation has been determined to be less than $13\%$.  Because we have nearly isotropic distribution (i.e., anisotropy that only results from statistical variations), the error should be relatively consistent throughout the coarsening process and therefore not impact our investigation of the evolution of the TPB length.

Not all of the TPBs are active; active TPBs must be simultaneously in contact with the three active phases that facilitate simultaneous transport of fuel, electrons, and oxygen ions. Since the electrochemical reactions occur only at or near these active TPBs, it is these active TPBs that contributes to the anode performance.  Before identifying the active TPBs, we must first identify the active phases.  Physically, a Ni phase cluster is active only if it connects the TPB sites to the current collector; a YSZ cluster is active only if it connects the TPBs to the electrolyte; and a pore cluster is active only if it connects the TPBs to the gas flow channels. However, the determination of such connectivities is feasible only if an anode structure that spans from the electrolyte to the current collector and to the gas channel is available. Our reconstructed microstructure, in contrast, constitutes only a portion of the entire span.  
  
As a compromise, we follow the procedures described in Ref.~\cite{JWilson}, which categorize each bulk region into active, dead-end, or isolated clusters.  Active clusters are clusters that are connected to at least two sides of the sample domain boundaries, while isolated clusters are those that are not connected to any of the sides and are therefore electrochemically inactive.  Dead-end clusters are defined as those that are in contact with only one side of the sample domain. In Ref.~\cite{JWilson}, a pre-smoothing procedure is used before this calculation to avoid artifacts resulting from image processing.  Because PFM simulations naturally smooth microstructures over the lengh scale of the interfacial thickness, the pre-smoothing procedure is not necessary in our analysis. Note that in this procedure, a voxel is assumed to belong to a cluster if at least one of the 6 nearest neighbors is in the cluster. The active TPBs are then identified as those simultaneously in contact with the active networks of all three different phases.  The TPB is marked as inactive as long as one of the phases adjacent to a TPB voxel is isolated.  The remainder of the TPBs are considered unknown in terms of activity.

\section{Results and Discussion}

\subsection{Phase connectivity and TPB activity}
\label{phaseCon}

TPB length has been recognized as one of the most important geometric parameter in a three-phase SOFC anode that influences the electrochemical performance. However, how the coarsening of microstructures affects the TPB activity has not been fully explored. As previously mentioned, TPBs are active only if they are simultaneously connected to conducting ionic, electronic, and gaseous transport pathways (phases). Therefore, the connectivity evolution of a phase induced by coarsening can alter the activity of TPBs and the effective conductivity of that phase. In addition, the evolution of the anode microstructure results in changes in the amounts and distributions of TPBs. In turn, coarsening has significant impacts on the overall performance of the electrode. By simulating coarsening with our models, a series of microstructures are acquired at various stage of evolution. Using the methods described in section~\ref{PconTPBact}, we can analyze the evolution of the active TPB length based on the evolving TPB distribution and phase connectivity during coarsening.  
\begin{figure}[htb!]
   \centerline{\includegraphics[width=9.cm,height=9.cm]{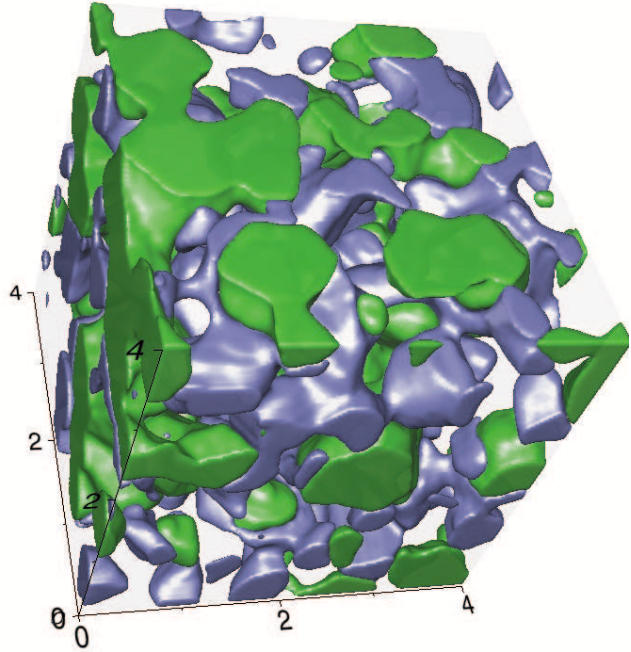}}
	\caption{Initial Ni-YSZ anode microstructure for a set of larger simulations. The dimensions of the microstructure are 4 $\mu$m $\times$ 4 $\mu$m $\times$ 4 $\mu$m. The Ni, pore, and YSZ phases are represented in green, blue, and semi-transparent, respectively (with volume fractions 23.8\%, 18.7\%, and 55.4\%, respectively).  The initial TPB density is $\sim$5.2$~\mu$m$^{-2}$. \label{fig:InitMS}}
\end{figure}

The initial microstructure of our simulation is based on an FIB-SEM reconstructed SOFC anode of dimensions 9.73 $\mu$m $\times$ 8.35 $\mu$m $\times$ 11.2 $\mu$m~\cite{James}.  We selected a 4.0 $\mu$m $\times$ 4.0 $\mu$m $\times$ 4.0 $\mu$m portion of the specimen (Fig.~\ref{fig:InitMS}), resolved by a domain of a 126$\times$126$\times$126 computational grid.  The governing evolution equations are solved using the finite-difference algorithm implemented using Open-MP. No-flux boundary conditions are imposed on the computational domain boundaries to reflect the assumption that materials that comprise the SOFC anode are conserved.

In the simulated sample, the volume fractions are 23.8\%, 18.7\%, and 57.5\% for the Ni, pore, and YSZ phases, respectively. Each phase region is categorized as an active, isolated, or dead-end cluster according to the procedure described in Sec.~\ref{PconTPBact}.  By examining the YSZ phase with the aforementioned nearest-neighbor scheme, we find that, in the initial sample, the entire YSZ phase is fully percolated and active within the volume. Using the same procedure, we find that there are 89.1\% active, 6.2\% isolated, and 4.7\% dead-end clusters within the Ni phase, and 94.7\% active, 1.9\% isolated, and 3.4\% dead-end clusters in the pore phase. 

Because our 3D microstructure data are represented as a set of voxels with values corresponding to various phases, this categorization procedure seems similar to the problem that consists of identifying the percolating clusters in a finite system on the basis of a simple cubic network with a coordination number of 6.  According to the percolation theory, the site percolation of a phase in a simple cubic network is achieved when the volume fraction of that phase is above a threshold value of 0.3116.  The YSZ phase in our sample is fully percolated and active because its volume fraction is much higher than this threshold value. However, it is surprising that nearly $90\%$ of the volume of the Ni and pore phases are active while their volume fractions are much lower than 0.3116.  This finding may be attributed to two reasons. First, the functional layer of our Ni-YSZ anode sample is fabricated by sintering a 50/50wt$\%$ NiO-YSZ mixture and exposing it to humidified H$_2$ to reduce NiO to Ni. This specific fabrication process results in highly percolated pore and Ni phases, even at volume fractions below 0.3116. Second, in our characterization algorithm, a cluster is considered active if it connects any two domain boundaries of the sample volume. Therefore, a cluster with a length scale smaller than the sample dimensions can be active as long as the cluster connects two neighboring domain boundaries of the sample. In terms of identifying the percolated clusters, this criterion is less stringent than the percolation theory that requires that the cluster is percolated over an infinite volume.    

\begin{figure}[htb!]
   \centerline{\includegraphics[width=11.06cm,height=7.6cm]{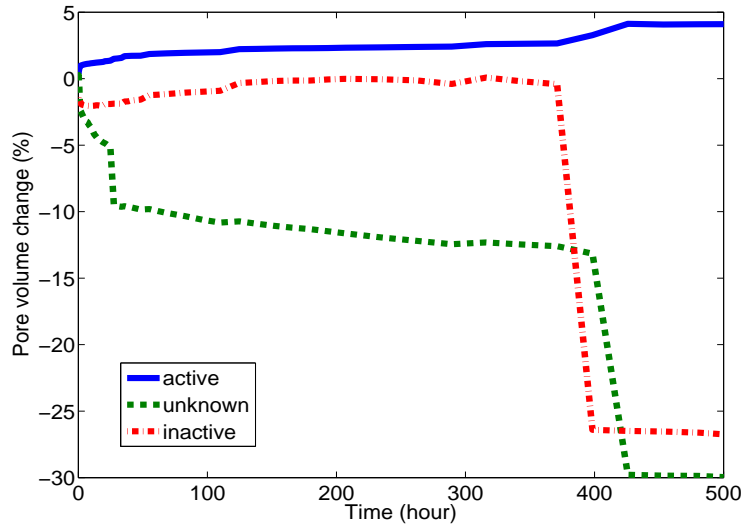}}
	\caption{The pore volume change in three different categories over 500 hours of coarsening. The change in each category is relative to its initial value. \label{fig:Pore_Vol_1000hr}}
\end{figure}

A crucial question arises is how these phase clusters evolve with coarsening.  The coarsening dynamics are greatly simplified in Model B because the Ni surface is the only mobile interface in the system and the YSZ phase is immobile. While coarsening progresses, the system reduces its total energy by reducing mobile regions with high curvatures (which possess smaller length scales and larger surface to volume ratios). Because the Ni and pore phases share the same Ni-pore interface, the mobile surfaces are the same for both phases.  The driving force for material transport thus depends on the curvature gradient, which is inversely linked to the characteristic length scale that is the inverse of the interfacial area per unit volume of the phase ($S_V^{-1}$). For example, the characteristic length scale of the Ni structure is defined by the volume of Ni divided by the total interfacial area of Ni. Note that this definition of the characteristic length scale is well suited for analyzing multiphase composites with unequal volume fractions because it takes into account the effects of volume on the typical size of a phase domain.  For example, if the volume fraction is large, then we expect that the typical size of the domain be larger even if the surface area is the same (this is the case in a two-phase system where that phase is the majority phase). Using this definition, an average Ni length scale 36\% larger than the average pore length scale is found in our sample. Inactive pore regions should thus evolve faster and possibly merge with active regions. 

According to the simulation results, during the first 500-hr simulation, the Ni volume fractions in each category and the mean Ni length scale remain roughly constant.  In contrast, as shown in Fig.~\ref{fig:Pore_Vol_1000hr}, the dead-end and isolated pore volumes decrease significantly, which confirms the faster evolution of regions associated with smaller length scales during coarsening.  Interestingly, the active-pore volume increases in our simulation due to the fact that the smaller isolated and dead-end clusters merge into the larger pore clusters, that are most likely in the active cluster category. One interesting observation is that even though clusters with higher curvatures possess higher free energy than those with lower curvatures, they may be trapped in a local equilibrium state because the transport pathway is confined to the Ni-pore interface. This kinetic constraint explains why the isolated and dead-end clusters in the pore phase did not suffer major loss of volume after 500 hours of coarsening.

\begin{figure}[htb!]
   \centerline{\includegraphics[width=11.06cm,height=7.6cm]{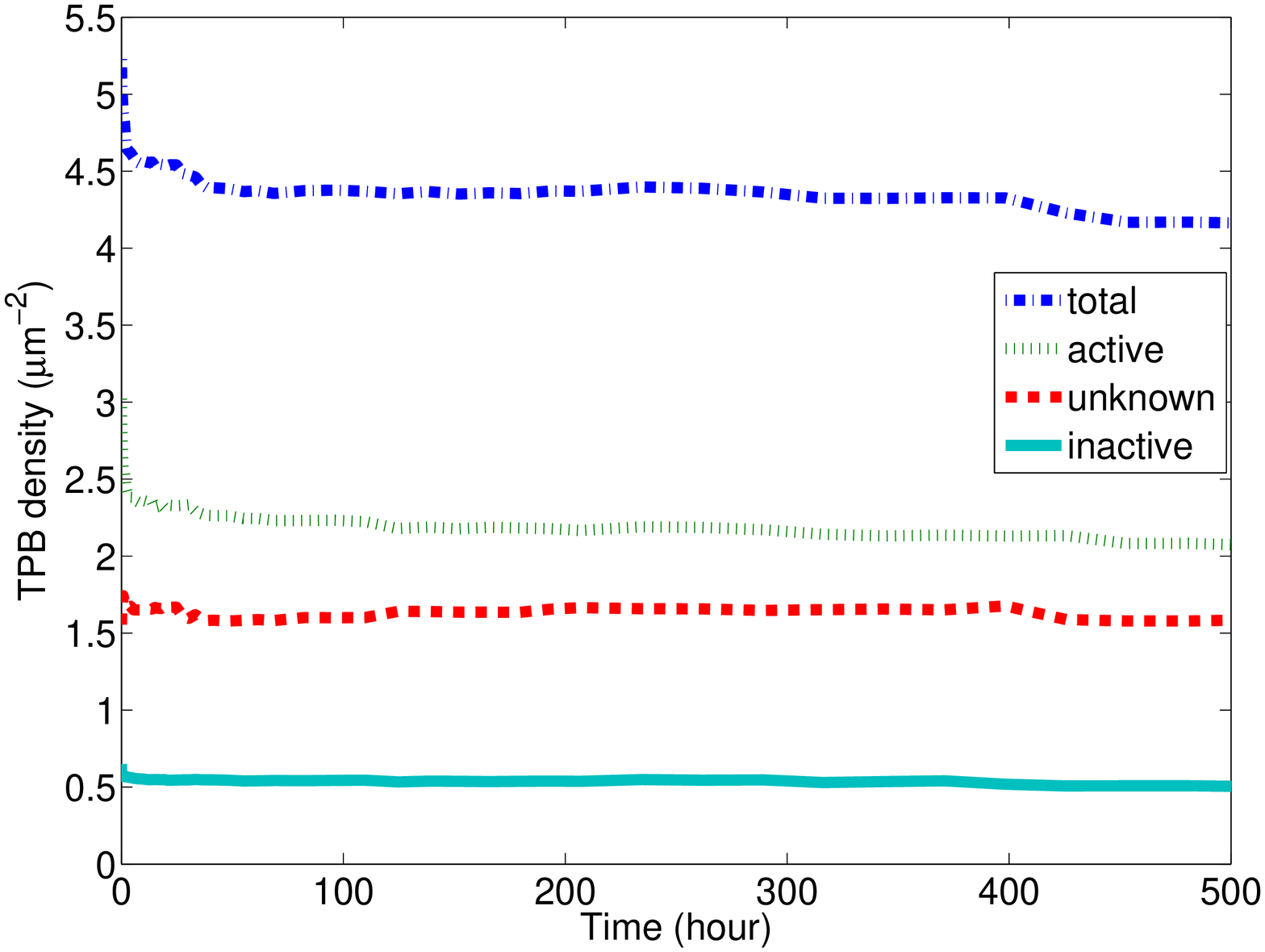}}
	\caption{TPB density change in each category over 500 hours of coarsening. The active TPB length reduction is $\sim$89$\%$ of the total TPB reduction after 500 hours. \label{fig:TPB_OT}}
\end{figure}

The evolution of TPBs is correlated with the evolution of the phases that comprise the SOFC anode.  The material transport among different clusters of each constituting phases not only changes the distribution of the TPBs but also dictates their activity.  As shown in Fig.~\ref{fig:TPB_OT}, the reduction trend of the active TPB density agrees well with the total TPB density. The overall TPB density decreased by 20.2\% after a 500-hour coarsening, while the active and isolated TPB density decreased by 31.4\% and 22.3\%, respectively. In contrast, the unknown TPB density slightly increased by 1.6\%. Most of the TPB density reduction comes from the 31.4\% decrease in the active TPBs. At the early stage of coarsening, the rapid reduction of active TPBs is due to the local coarsening of high-curvature microstructural features. The small increase in the unknown TPBs observed in the early stage of evolution stems from coalescence of inactive Ni domains, which occurs rarely. Since the evolution of TPBs is sensitive to local microstructural features, it is only indirectly correlated to the coarsening kinetics.   

To summarize, our findings indicate that the coarsening of a Ni-YSZ anode consumes the mobile phase clusters with high curvatures.  The isolated and dead-end pore clusters are thus the first clusters that evolve, leading to coarsening of the pore phase, which is indicated by the change of $S_{V(pore)}^{-1} = 0.1105~\mu$m to $S_{V(pore)}^{-1} = 0.1154~\mu$m, i.e., a $4.4\%$ increase over 500 hours.  While the Ni phase has the same mobile area as the pore phase, its mean characteristic length scale, which is defined by $S_{V(Ni)}^{-1}$, is significantly larger than that of the pore phase and results in a slower evolution. In addition, the Ni-pore area reduction is found to be balanced by the increase of the Ni-YSZ interfacial area.  The coarsening has thus no significant effect on the mean length scale of Ni, which varies from $S_{V(Ni)}^{-1} = 0.1711~\mu$m to $S_{V(Ni)}^{-1} = 0.1719~\mu$m (which corresponds to a $0.5\%$ change during the course of our simulation.  In turn, the coarsening leads to a significant reduction in TPBs and the active TPB sites decreases in a very similar fashion as the overall TPBs. However, the TPB evolution is sensitive to microstructural details and is only weakly linked to the evolution of bulk phases during coarsening. 

\subsection{Tortuosity of pores}

\begin{figure}[htb!]
   \centerline{\includegraphics[width=11.06cm,height=7.6cm]{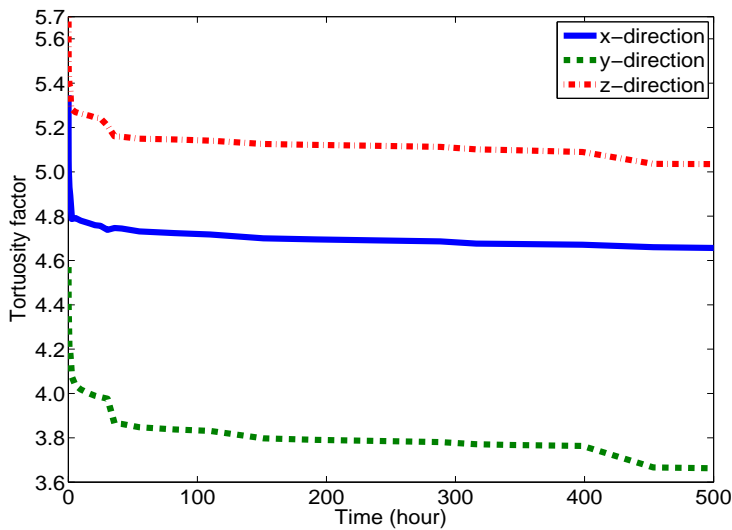}}
	\caption{Tortuosity factors in three orthogonal directions of the pore phase over 500 hours of coarsening. \label{fig:TortuRedRate}}
\end{figure}

Tortuosity represents the geometric aspect of the transport property of a phase that comprises a composite.  Therefore, the tortuosity of the pore phase plays a crucial role in the generation of electricity because the fuels need to be transported through the pores to reach the reaction sites.  The tortuosity factors of the pore phase are evaluated in three orthogonal directions over a 500-hr coarsening period.  As shown in Fig.~\ref{fig:TortuRedRate}, although the absolute values differ, the tortuosity factors decrease moderately in all directions during coarsening.  This trend is consistent with the fact that the active pore volume increases slightly during coarsening (see Sec.~\ref{phaseCon}, Fig.~\ref{fig:Pore_Vol_1000hr}).  These results indicate a coalescence of pore clusters and a lack of breakup of active clusters, which lead to somewhat enhanced transport properties. 

\subsection{The active Ni surface}

One advantage of SOFCs is their ability to operate with hydrocarbon fuels as well as hydrogen. This is because hydrocarbon fuels can be internally reformed to hydrogen with the catalytic effect of Ni at the typical operating temperatures of SOFCs.  This reforming rate is highly related to the active Ni surface area.  Depending on the fuel, the reforming kinetics can be very different and involve multiple elementary reaction steps; however, the reforming reactions of all types of fuels involve gas species and electron transfer, regardless of the detailed reforming mechanisms.  Therefore, the active Ni surface for hydrocarbon reforming must reside at the interface of active pore clusters and active Ni clusters.

\begin{figure}[htb!]
   \centerline{\includegraphics[width=11.06cm,height=7.6cm]{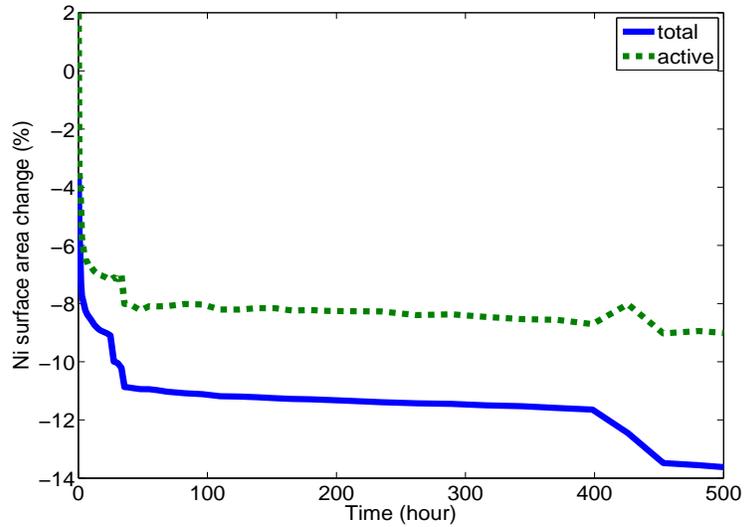}}
	\caption{Ni surface area reduction over 500 hours of coarsening. \label{fig:NiSARedRate}}
\end{figure}

To evaluate the catalytic performance of the Ni phase in SOFC anodes, an important question to ask is how coarsening affects the active Ni surface area.  The evolution of an active Ni surface area is dictated by the evolution of active clusters of Ni and pore phases.  The active Ni surface is identified as the interfaces between active Ni and active pore clusters.  As shown in Fig.~\ref{fig:NiSARedRate}, the active Ni surface decrease by 9\% after 500-hr coarsening, while the total Ni-pore interface decrease by 13.6\%. Our finding indicates that coarsening reduces more inactive Ni surfaces than those in contact with the active clusters of Ni and pore.  This is due to the fact that smaller clusters, which have smaller length scales and thus have larger driving force toward coarsening, coarsen faster than the larger ones.

\subsection{Size effects and evolution kinetics}
\label{SizeEff}

We next simulate coarsening in Ni-YSZ anode samples of different sizes, namely, (3.2~$\mu$m)$^3$, (4.0~$\mu$m)$^3$, and (4.8~$\mu$m)$^3$, to examine the size effects. Each of these samples is acquired from a portion of the original 914.55~$\mu$m$^3$ sample.  Specifically, the sample of dimensions (3.2~$\mu$m)$^3$ is cropped from the sample of dimensions (4.8~$\mu$m)$^3$, while the sample of dimensions (4.0~$\mu$m)$^3$ belongs to a different portion of the original sample. 

\begin{figure}[htb!]
   \centerline{\includegraphics[width=11.06cm,height=7.6cm]{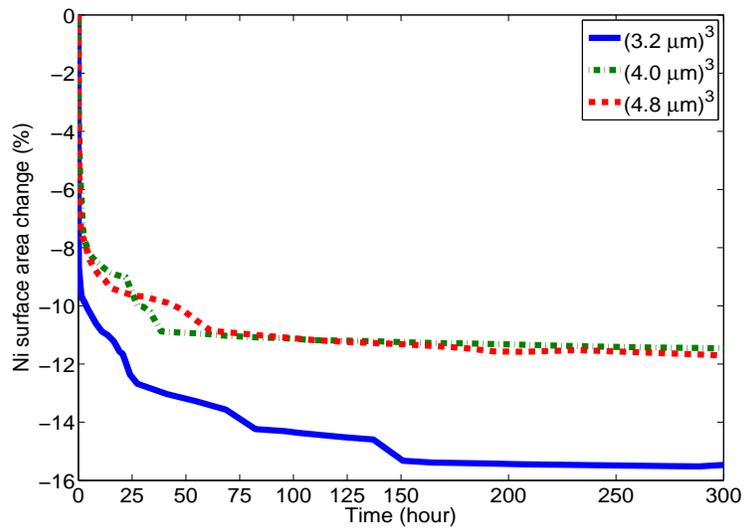}}
	\caption{Reduction of the Ni surface areas for three different sample sizes over 300 hours of coarsening. \label{fig:Size_NiSurfA}}
\end{figure}

\begin{figure}[htb!]
   \centerline{\includegraphics[width=11.06cm,height=7.6cm]{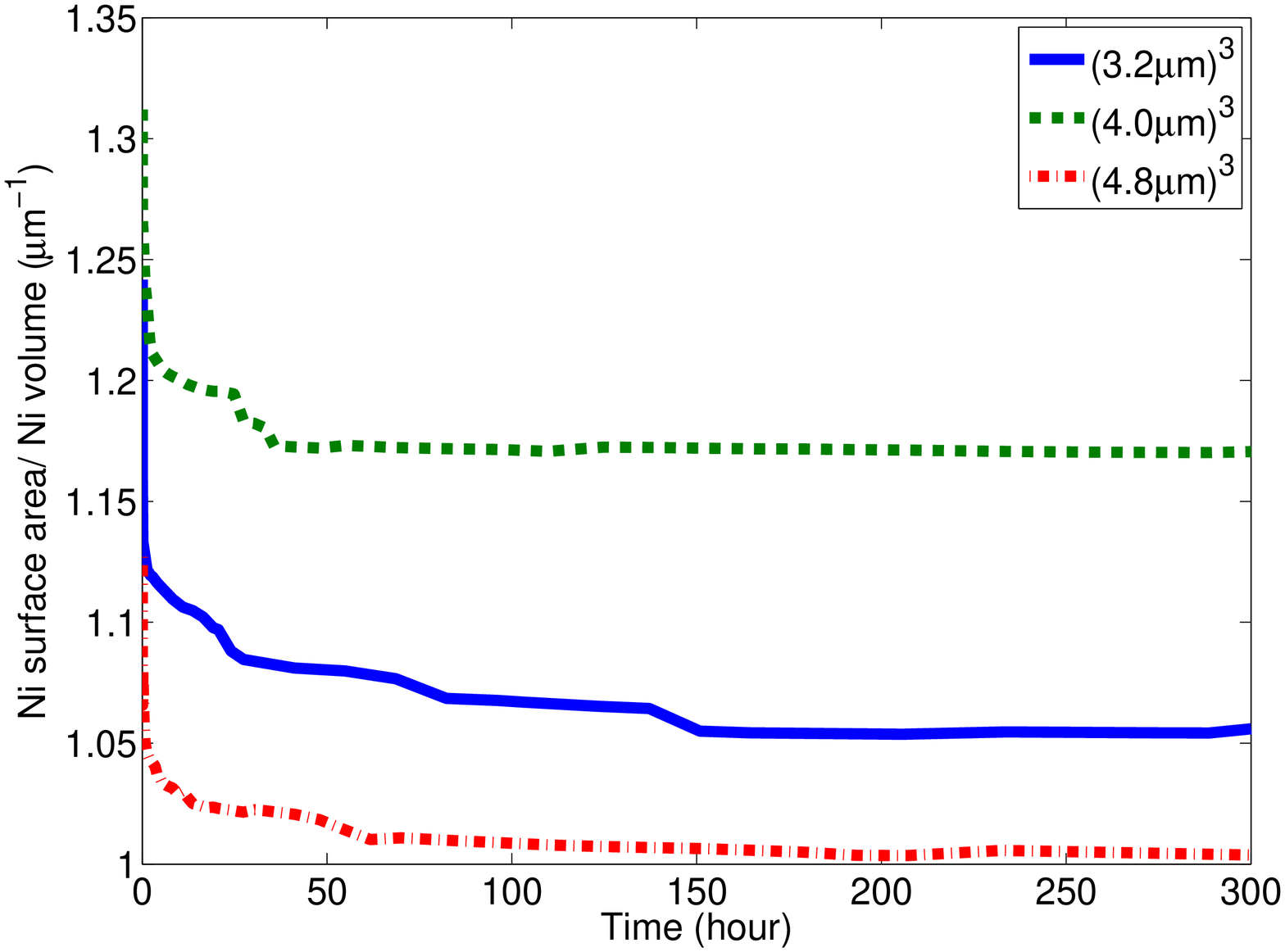}}
	\caption{Reduction of the Ni surface area per Ni volume for three different sample sizes over 300 hours of coarsening. \label{fig:Size_NiS_v}}
\end{figure}

The Ni surface areas and the TPB lengths are compared over the first 300 hours of the coarsening period. The evolution kinetics can be best interpreted from the evolution of the Ni surface area in our Model B simulations. As shown in Figs.~\ref{fig:Size_NiSurfA} and \ref{fig:Size_NiS_v}, the Ni surface area reduction rate of the (4.0~$\mu$m)$^3$ and the (4.8~$\mu$m)$^3$ specimen are very similiar, while the reduction rate of the (3.2~$\mu$m)$^3$ specimen deviates from the other two significantly even though the specimen is a portion of the (4.8~$\mu$m)$^3$ sample. In addition, some reduction steps appear in the (3.2~$\mu$m)$^3$ curve whereas the other two curves are relatively smooth. This suggests that the two larger sample sizes may be sufficient to eliminate the boundary effects from the simulations, while smaller volumes would likely suffer from them and may not contain enough particles or statistics for coarsening simulations.

\begin{figure}[htb!]
\centerline{\includegraphics[width=11.06cm,height=7.6cm]{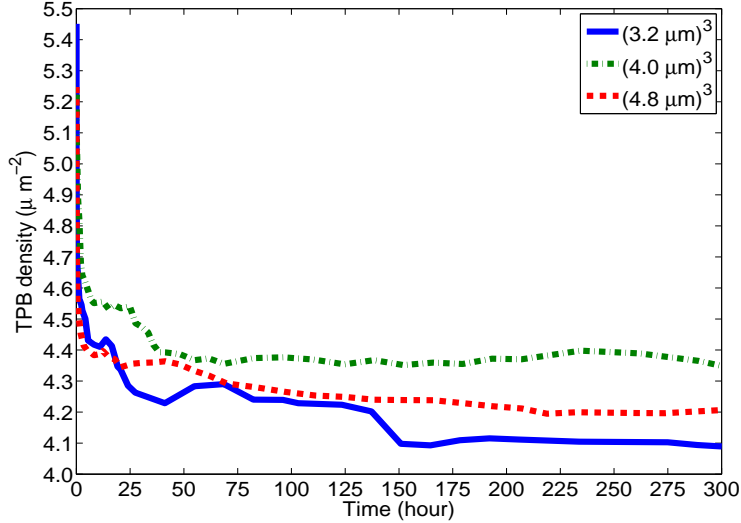}}
	\caption{TPB densities for three differnt sample sizes over 500 hours of coarsening. The initial TPB densities of the (3.2 $\mu$m)$^3$, (4 $\mu$m)$^3$, and the (4.8 $\mu$m)$^3$ samples are 5.45 $\mu$m$^{-2}$, 5.22 $\mu$m$^{-2}$, and 5.25 $\mu$m$^{-2}$, respectively. \label{fig:Size_TPB}}
\end{figure}

As shown in Fig.~\ref{fig:Size_TPB}, the change of the TPB densities of the three specimens are compared over 300-hr coarsening.  Although the behavior of the TPBs is only indirectly related to the evolution kinetics, we find similar trend in the TPB evolution among the three specimens as those found in the Ni surface area.  Although the TPB density reduction kinetics are different, after 300-hour coarsening, the difference of the TPB densities between the two larger samples is less than 3\%. Since TPB evolution is more sensitive to microstructures, this difference may serve as the indication of the local variation due to microsctructural details. Thus, in order to determine whether the two larger sample volumes are sufficient to represent the microstructure of the anode, we would need to simulate a larger sample volume.
 
Although rapid reductions of TPBs and the Ni surface area at early stages of coarsening is a result of the large thermodynamic driving force for mass transport, which mostly results from regions with high curvatures, the overly steep slopes in Figs.~\ref{fig:Size_NiSurfA} and~\ref{fig:Size_TPB} deserve further investigations.  

\begin{figure}[htb!]  \centerline{\includegraphics[width=10.8cm,height=10.8cm]{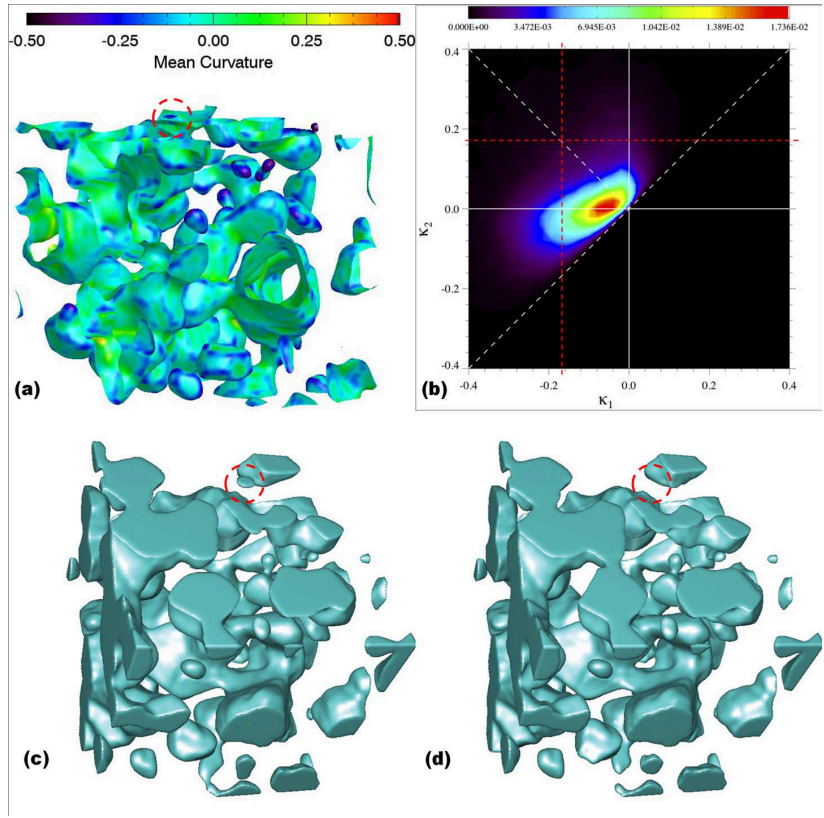}}
	\caption{Numerical smoothing of Ni microstructures in Model B. The circled particle is numerically smoothed after a very short time of coarsening. (a) mean-curvature ($H = (\kappa_1+\kappa_2)/2$) plot of Ni interfaces; (b) ISD diagram of Ni interfaces; (c) the Ni microstructure before coarsening; (d) Ni microstructure after coarsening for 3 minutes. (Figures (a) and (b) were plotted by Chal Park in Thornton's group.) \label{fig:UnderResolve}}
\end{figure}

To examine the length scales involved in the microstructures, the mean curvature and the interfacial shape distribution (ISD) of the Ni interfaces in the (4 $\mu$m)$^3$ specimen are plotted in Fig.~\ref{fig:UnderResolve} (a) and (b), respectively. Due to the diffuse interface nature of our models and the model parameters we utilized in our simulations (based on the resolution that made simulations feasible), very small microstructural features with absolute principal curvature values larger than 0.17 (which accounts for approximately $30\%$ of the Ni interfacial area) are not fully resolved. As shown in Fig.~\ref{fig:UnderResolve} (c) and (d), if those numerically under-resolved regions are mobile, they will be smoothed out very rapidly, leading to coarsening that is faster than expected.

This numerical smoothing due to under-resolution contributes in part to the rapid evolution in the early stage of coarsening simulations and leads to an overestimation of the TPB reduction rate.  However, these mobile, high-curvature regions inherently possess very large thermodynamic driving force and will coarsen sooner or later.  Therefore, even though the rapid evolution at the early stage of simulations may be caused by the insufficient resolution of microstructures, the stabilized value of TPB density after further coarsening remains unaffected by it. That is, the prediction of the stabilization of TPBs, as well as its predicted value, remains robust, even though the early kinetics may be overestimated. Since the microstructures of Ni-YSZ anodes are found to stabilize after coarsening for a period of time, the final or stablized TPB density rather than the short-term evolution kinetics is of importance for the operation of SOFCs.

\section{Conclusion}
We developed two models to study the coarsening kinetics in the Ni-YSZ anode.  An asymptotic analysis was conducted to link our simulation results to the physical system.  The size effects were studied and an error analysis was performed to validate our models.  For the model parameters selected, no obvious boundary effects were observed when the simulated domain was larger than (4~$\mu$m)$^3$, even though the sample size was still insufficient to be statistically representative of the entire microstructure at this volume.  In addition, the use of the selected mesh resolution had very minor effects on the quasi-equilibrium contact angles and evolution kinetics. While the short-term evolution kinetics may be affected by the insufficient resolution of microstructures, the amount of TPBs and other properties after long-term coarsening can be identified by our simulations. 

Although our models contain approximations and the simulation results may be affected by uncertainties in the material properties, a reasonable agreement could be found for the TPB length reduction between coarsening experiments and simulation results using Model B.  Unlike Model B, Model A overestimates the TPB reduction due to the evolution of the YSZ structure that is induced by an excess mobility at triple junctions built into the model.  Our simulations show that the major portion of TPB reduction occurs during the early stages of coarsening and that stability of TPBs is observed provided that YSZ is nearly immobile. 

The evolution of active TPBs, active Ni surface areas, and tortuosity of the pore phase are investigated in our simulations.  We found that coarsening has a smaller impact on the active Ni surface areas than on their total respective amounts or their inactive counterparts.  In addition, the tortuosity of the pore phase was found to decrease slowly in all three orthogonal directions during coarsening.  These phenomena are due to the fact that percolated phase clusters that dictate the active parameters and transport have typically larger characteristic length scales, $S_V^{-1}$, than isolated/dead-end phase clusters. Smaller clusters experience larger driving forces for coarsening, while larger clusters may grow at the expense of these regions if transport can be facilitated between them.  Active Ni surface areas are also less reduced than their dead/isolated counterparts, and the transport property of the pore phase is slightly enhanced.  In contrast, the reduction of active TPBs are found to account for most of the total TPBs reduction, which suggests that the TPB evolution is sensitive to microstructural details and is indirectly related to the evolution of bulk phases during coarsening. 
   
The proposed coarsening models provide insights into directing the design of anode microstructures.  The model framework is general and can be applied to many other three-phase coarsening systems. Further experiments can help validate and improve these models.

\section{Appendix}
\subsection{Asymptotic analysis}
The asymptotic analysis that was used is similar to that described in~\cite{DongHee}. Briefly, we first determine the outer solution. Substituting equation \ref{eq:NDmu} into \ref{eq:NDCH3}, dropping the overbars, and replacing the variables $\delta$ with $\zeta$, $\mu$ with $\mu_{out}(\overrightarrow{r},t,\zeta)$, and $\phi$ with $\phi_{out}(\overrightarrow{r},t,\zeta)$, we obtain the non-dimensional evolution equations
\begin{equation}
    \zeta^{2} \frac{\partial \phi_{out}}{\partial t} = \triangledown \cdot \left [ M(\phi_{out}) \triangledown \left ( \frac{\partial f}{\partial \phi_{out}} - \zeta^{2} \triangledown^{2} \phi_{out} \right ) \right ]
	\label{eq:NDCH4}
\end{equation}
\begin{equation}
    \mu_{out} = \frac{\partial f}{\partial \phi_{out}} - \zeta^{2} \triangledown^{2} \phi_{out}
	\label{eq:MuOut}
\end{equation}

\subsubsection{Outer expansion}
An outer expansion is performed by expanding the field $\phi_{out}$ and $\mu_{out}$ in our model in powers of $\zeta$ as follows:
\begin{equation}
    \phi_{out}(\overrightarrow{r},t,\zeta) = \phi_{out}^{(0)}(\overrightarrow{r},t) + \phi_{out}^{(1)}(\overrightarrow{r},t) \zeta + \phi_{out}^{(2)}(\overrightarrow{r},t) \zeta^{2} + \cdots
	\label{eq:Cexp}
\end{equation}
\begin{equation}
    \mu_{out}(\overrightarrow{r},t,\zeta) = \mu_{out}^{(0)}(\overrightarrow{r},t) + \mu_{out}^{(1)}(\overrightarrow{r},t) \zeta + \mu_{out}^{(2)}(\overrightarrow{r},t) \zeta^{2} + \cdots
	\label{eq:Muexp}
\end{equation}

Substituting Eq.~\ref{eq:Cexp} into Eq.~\ref{eq:NDCH4}, we obtain to the zeroth order in $\zeta$
\begin{equation}
    \triangledown \cdot \left [M(\phi_{out}^{(0)}) \triangledown \left (\frac{\partial f}{\partial \phi_{out}^{(0)}} \right ) \right ] = 0.
    \label{eq:Outexp}
\end{equation}
Eq.~\ref{eq:Outexp} is satisfied with the solution of 
\begin{eqnarray}
    \phi_{out}^{(0)}(\overrightarrow{r})=1~~~~~~ \forall \overrightarrow{r} \in \Omega_{+} \nonumber \\
    \phi_{out}^{(0)}(\overrightarrow{r})=0~~~~~~ \forall \overrightarrow{r} \in \Omega_{-},
\label{eq:OutSolu}
\end{eqnarray}
where $\Omega_{\pm}$ are the two bulk phase regions.  This solution asserts that far from the interface, we have one of the equilibrium phases. By substituting Eqs.~\ref{eq:Cexp} and~\ref{eq:Muexp} into Eq.~\ref{eq:MuOut} and collecting the zeroth order of $\zeta$, we find
\begin{equation}
    \mu_{out}^{(0)} = \frac{\partial f}{\partial \phi_{out}^{(0)}}.
    \label{eq:MuC0}
\end{equation}
Substituting Eq.~\ref{eq:OutSolu} into Eq.~\ref{eq:MuC0} yields 
\begin{equation}
    \mu_{out}^{(0)}(\overrightarrow{r} \in \Omega_{\pm}) = 0.
    \label{eq:MuC0solu}
\end{equation}

\subsubsection{Inner expansion}
To denote the inner solution, we replace the variable $\delta$ with $\zeta$, $\mu$ with $\mu_{in}$ and $\phi$ with $\phi_{in}$. To facilitate the inner expansion, we introduce a moving coordinate system (M.C.S.): one coordinate $r$ is parallel to $\triangledown \phi$, while the other coordinate is the arclength $s$ along the interface (which is located at $r = 0$).  The operators close to the interface ($r \approx 0$) are given by:
\begin{equation}
    \triangledown = \overrightarrow{e}_{r} \frac{\partial}{\partial r} + \overrightarrow{e}_{s} \frac{\partial}{\partial s},~~~~ \triangledown^{2} = \frac{\partial^{2}}{\partial r^{2}} + \kappa_{c} \frac{\partial}{\partial r} + \frac{\partial^{2}}{\partial s^{2}},
    \label{eq:Opt}
\end{equation}
where $\kappa_c$ is the local curvature.  Introducing a stretched variable $z=r/\zeta$,
\begin{equation}
    \triangledown = \overrightarrow{e}_{r} \frac{1}{\delta} \frac{\partial}{\partial z} + \overrightarrow{e}_{s} \frac{\partial}{\partial s},~~~~ \triangledown^{2} = \frac{1}{\delta^{2}}\frac{\partial^{2}}{\partial z^{2}} + \kappa_{c} \frac{1}{\delta} \frac{\partial}{\partial z} + \frac{\partial^{2}}{\partial s^{2}}.
    \label{eq:StOpt1}
\end{equation}
\begin{equation}
    \frac{\partial}{\partial t} = \left ( \frac{D}{D t} \right ) - V_{n} \frac{1}{\delta} \frac{\partial}{\partial z} - V_{s} \frac{\partial}{\partial s},
    \label{eq:StOpt2}
\end{equation}
where $V_{n}$ and $V_{s}$ are the normal and tangential velocities, and $D/Dt$ represents the material derivative, which is the derivative taken based on the M.C.S.  Multiplying Eq.~\ref{eq:NDCH4} by $\zeta^{2}$ and rewriting it in terms of a new field variable $\phi$ in the M.C.S. yields
\begin{equation}
  -V_{n} \frac{\partial \phi}{\partial z} \zeta^{3} + \left (\frac{D \phi}{D t} - V_{s} \frac{\partial \phi}{\partial s} \right ) \zeta^{4} = \frac{\partial}{\partial z} \left ( M \frac{\partial \mu}{\partial z} \right ) + \zeta \kappa_{c} M \frac{\partial \mu}{\partial z} + \zeta^{2} \frac{\partial}{\partial s} \left (M \frac{\partial \mu}{\partial s} \right ).
  \label{eq:InExpGov}
\end{equation}
Expanding the fields in the power of $\zeta$ gives:
\begin{equation}
    \phi_{in}(z,s,t,\zeta) = \phi_{in}^{(0)}(z,s,t) + \phi_{in}^{(1)}(z,s,t) \zeta + \phi_{in}^{(2)}(z,s,t) \zeta^{2} + ...
	\label{eq:PhiExp}
\end{equation}
\begin{equation}
    \mu_{in} = \mu_{in}^{(0)} + \mu_{in}^{(1)} \zeta + \mu_{in}^{(2)} \zeta^{2} + ...
	\label{eq:MuExp}
\end{equation}
Substituting Eq.~\ref{eq:PhiExp} and~\ref{eq:MuExp} into Eq.~\ref{eq:InExpGov} and collecting the zeroth order of $\zeta$ leads to:
\begin{equation}
    \frac{\partial}{\partial z} \left [ M \left (\phi_{in}^{(0)} \right ) \frac{\partial{\mu_{in}^{(0)}}}{\partial z} \right ] = 0.
    \label{eq:InExp0}
\end{equation}
Integrating Eq.~\ref{eq:InExp0} with respect to $z$ yields
\begin{equation}
     M (\phi_{in}^{(0)} ) \frac{\partial{\mu_{in}^{(0)}}}{\partial z} = g_{0}(s,t).
    \label{eq:IntIn0}
\end{equation}
Taking the limit $z \rightarrow \pm \infty$ and matching with Eq.~\ref{eq:MuC0solu}, we obtain
\begin{equation}
     M (\phi_{in}^{(0)} ) \frac{\partial{\mu_{in}^{(0)}}}{\partial z} = 0.
    \label{eq:MatchIn0}
\end{equation}
Because $\phi_{in}^{(0)}$ should vary smoothly from 1 to 0 as $z$ transitions from $+ \infty$ to $- \infty$, $M(\phi_{in}^{(0)} )$ is expected to be nonzero. We thus have 
\begin{equation}
     \frac{\partial{\mu_{in}^{(0)}}}{\partial z} = 0 \quad \mathrm{or} \quad \mu_{in}^{(0)}(z,s,t) = g_{1}(s,t).
    \label{eq:MuIn0od}
\end{equation}
Matching Eq.~\ref{eq:MuIn0od} with the zeroth order in outer field, the profile of $\phi_{in}^{0}$ along the $z$-direction is governed by \begin{equation}
    \mu_{in}^{(0)} = 0 = \frac{\partial f}{\partial \phi_{in}^{(0)}} - \frac{\partial^{2} \phi_{in}^{(0)}}{\partial z^{2}},
\end{equation}
which leads to the following relationship that is useful in change of veriables:
\begin{equation}
    \frac{\partial \phi_{in}^{(0)}}{\partial z} = \sqrt{2 f(\phi_{in}^{(0)})}.
    \label{eq:ChVar}
\end{equation}
Collecting the first order of $\zeta$ in Eq.~\ref{eq:InExpGov} gives:
\begin{equation}
    \frac{\partial}{\partial z} \left (\frac{\partial \mu_{in}^{(1)}}{\partial z} \right ) + \kappa_{c} \frac{\partial \mu_{in}^{(0)}}{\partial z} = 0,
\end{equation}
which in turn gives $\mu_{in}^{(1)} = \mu_{in}^{(1)}(s)$, or $\mu_{in}^{(1)}$ is independent of $z$.  Similarly, collecting the second order of $\zeta$ in Eq.~\ref{eq:InExpGov}, we find that $\mu_{in}^{(2)} = \mu_{in}^{(2)}(s)$, or $\mu_{in}^{(2)}$ is independent of $z$.
Finally, by collecting the third order of $\zeta$ in Eq.~\ref{eq:InExpGov} and considering that $\mu_{in}^{(1)}$ and $\mu_{in}^{(2)}$ are independent of $z$, we have:
\begin{equation}
    -V_{n} \frac{\partial \phi_{in}^{(0)}}{\partial z} = \frac{\partial}{\partial z} \left [M(\phi_{in}^{(0)}) \frac{\partial \mu_{in}^{(3)}}{\partial z} \right ] + \frac{\partial}{\partial s} \left [M(\phi_{in}^{(0)}) \frac{\partial \mu_{in}^{(1)}}{\partial s} \right ].
    \label{eq:InO3}
\end{equation}
Integrating Eq.~\ref{eq:InO3} from $z = -\infty$ to $z = \infty$ gives:
\begin{equation}
    V_{n} =  - \frac{\partial}{\partial s} \left [\int\limits_{-\infty}^{\infty}M(\phi_{in}^{(0)}) dz \right ] \frac{\partial \mu_{in}^{(1)}}{\partial s} = - I \cdot \frac{\partial^{2} \mu_{in}^{(1)}}{\partial s^{2}},
    \label{eq:InIntO3}
\end{equation}
where we assume a constant $I = \int\limits_{-\infty}^{\infty} M(\phi_{in}^{(0)}) dz$ as it is independent of $s$.  Also, substituting Eq.~\ref{eq:PhiExp} and~\ref{eq:MuExp} into Eq.~\ref{eq:NDmu} and collecting the first order terms of $\zeta$ leads to:
\begin{equation}
    \mu_{in}^{(1)} = \frac{\partial^{2} f}{\partial \phi^{2}} \phi_{in}^{(1)} - \frac{\partial^{2} \phi_{in}^{(1)}}{\partial z^{2}} - \kappa_{c} \frac{\partial \phi_{in}^{(0)}}{\partial z}.
    \label{eq:Che0}
\end{equation}
Multiplying Eq.~\ref{eq:Che0} by $\partial \phi_{in}^{(0)}/\partial z$ and integrating over z from $-\infty$ to $\infty$ gives:
\begin{equation}
    \mu_{in}^{(1)} = - \kappa_{c} \int\limits_{-\infty}^{\infty} \left (\frac{\partial \phi_{in}^{(0)}}{\partial z} \right )^{2} dz.
    \label{eq:Che1}
\end{equation}
Performing a change of variables in Eq.~\ref{eq:Che1} using the relationship given in Eq.~\ref{eq:ChVar} leads to:
\begin{equation}
    \mu_{in}^{(1)} = - \kappa_{c} \int\limits_{0}^{1} \sqrt{2 f(\phi_{in}^{(0)})} d\phi = - \kappa_{c} \cdot J,
    \label{eq:Che2}
\end{equation}
, wherein we assume a constant $J = \int\limits_{0}^{1} \sqrt{2 f(\phi_{in}^{(0)})} d\phi_{in}^{(0)}$.  Combining Eq.~\ref{eq:Che2} and Eq.~\ref{eq:InIntO3} results in:
\begin{equation}
    V_{n} = I \cdot J \cdot \frac{\partial^{2} \kappa_{c}}{\partial s^{2}}
    \label{eq:Vn1}
\end{equation}
Performing the integration of $I$ and $J$ (dropping the superscripts and subscripts for convenience) leads to:
\begin{equation}
    J = \int\limits_{0}^{1} \phi (1-\phi) d\phi = \frac{1}{6}.
\end{equation}
\begin{equation}
    I = \int\limits_{-\infty}^{\infty} M(\phi) dz = \int\limits_{0}^{1} \frac{M(\phi)}{\sqrt{2f}} d \phi= \bar{M}_{s} \int\limits_{0}^{1} \phi (1-\phi) d\phi = \frac{\bar{M}_{s}}{6}.
\end{equation}
Substituting $I$ and $J$ into Eq.~\ref{eq:Vn1} and restoring the dimensions gives:
\begin{equation}
    V_{n} \frac{L^{3}}{D} = - \frac{\bar{M_{s}} L^{3}}{36} \frac{\partial^{2} \kappa_{c}}{\partial s^{2}}.
    \label{eq:Vn2}
\end{equation}

\bibliographystyle{unsrt}
\bibliography{SOFCanode}
\end{document}